\documentstyle[aps,prd]{revtex}

\input epsf.sty

\preprint{UCTP-107-99}

\title{Universality and the magnetic catalysis of chiral symmetry 
breaking}

\author{G. W.~Semenoff\thanks{On leave from 
        Department of Physics and Astronomy, 
        University of British Columbia, 
        6224 Agricultural Road, Vancouver, 
        British Columbia V6T 1Z1, Canada.}} 

\address{The Niels Bohr Institute, Blegdamsvej 17,
DK-2100, Copenhagen {\O} Denmark }

\author{I.A.~Shovkovy\thanks{
        On leave of absence from 
        Bogolyubov Institute for Theoretical Physics, 
        Kiev 252143, Ukraine.}
and L.C.R.~Wijewardhana}

\address{Physics Department, 
         University of Cincinnati,
         Cincinnati, Ohio 45221-0011}

\date{\today}

\draft
\begin{document}
\maketitle

\begin{abstract}
The hypothesis that the magnetic catalysis of chiral symmetry
breaking is due to interactions of massless fermions in their lowest
Landau level is examined in the context of chirally symmetric models
with short ranged interactions. It is argued that, when the
magnetic field is sufficiently large, even an infinitesimal
attractive interaction in the appropriate channel will break chiral
symmetry. 
\end{abstract}



\section{Introduction}

Field theoretical models in external electromagnetic fields are of
great interest and have attracted a lot of attention
\cite{JackK,Git,QHE,EW-mag}. They are relevant to the study of many
physical systems whose properties depend on  the effects produced by
external fields, especially when such effects are  non-perturbative
in nature. A particular example of such a situation is the so-called
magnetic catalysis of chiral symmetry breaking
\cite{GMS,Ng,Hong,Other,Vic,GMS-114} (see also \cite{Kle,Klim,Schr} 
for some earlier studies) which has many potential  applications in
condensed matter physics \cite{ssw,mav,cond} and in studies of the
early Universe \cite{univ}. 

It is well known that a sufficiently strong attractive interaction
between massless fermions  can result in a chiral symmetry breaking
condensate  and a dynamically generated fermion mass. In the
Nambu-Jona-Lasinio (NJL) model, for example, this occurs when the
interaction strength is at a particular large critical value. It is
also now known that the presence of an external magnetic field can
be a strong catalyst for this effect, leading in some cases to the
generation of a dynamical mass  for fermions even by very weak
attractive interactions. For example, in an external magnetic field,
the critical value of the NJL coupling is reduced to zero.

It was suggested in \cite{GMS} that this magnetic catalysis of
chiral symmetry breaking is a rather universal phenomenon and
that its main features are model independent. The arguments were
based  on the observation that the effect is primarily due to
the dynamics of the fermions from the lowest Landau level
which are gapless and should therefore dominate the behavior of
the system at long wavelengths. 

In this paper we will develop this idea further. We shall
investigate the low-energy dynamics of chirally  symmetric models
with a short range interaction in an external magnetic  field. The
idea is to construct an effective action  for the low energy degrees
of freedom. In strong external magnetic  fields, these are the
modes of charged particles in the lowest Landau level. In the absence
of  interactions, their effective dynamics is described by a field
theory whose dimensionality is two less than the spacetime
dimension. We shall take the hypothesis that, at least for strong
magnetic fields and short ranged  interactions, the dynamics of
chiral symmetry breaking can be understood  in the lower dimensional
effective theory.

As an illustration, we shall argue that this assumption is very
natural in 2+1 dimensions. There, the modes of the lowest Landau
level are non-propagating, and in the absence of interactions, the
ground state is highly degenerate and contains both chirally
symmetric and chirally non-symmetric states. Interactions, even with
infinitesimally small coupling constants, will split this degeneracy,
and whether chiral symmetry breaking occurs or not depends on
whether the interaction favors the chirally non-symmetric or chirally
symmetric ground states. One by-product of our analysis will be the
fact that the dynamical generation of parity violating masses seems
to be suppressed by magnetic fields. 

In the more complicated case of (3+1)-dimensional gauge theory, we
shall argue that the infrared dynamics is governed by an  effective
(1+1)-dimensional Gross-Neveu like theory with an infinite number of
flavors of fermions. We derive the most general low energy effective
action which is compatible with the symmetries of the four
dimensional theory. We then study the resulting  model using the
renormalization group. We will present a plausible argument for our
main conclusion:  that the presence of chiral symmetry breaking
depends crucially on the sign (but not the  magnitude) of one
particular moment of the coupling constants which we denote by
$g_v$. If it is of the appropriate sign, its infrared
renormalization group flow is to strong coupling and chiral symmetry
is broken. With the other sign it flows to  zero coupling and chiral
symmetry remains unbroken.

\section{(2+1)-dimensional paradigm}

Consider a system of relativistic (2+1)-dimensional fermions with 
the Dirac Hamiltonian
\begin{equation}
H_0=\left( \matrix{ i\vec\sigma\cdot\vec D & 0\cr 0 & 
-i\vec\sigma\cdot\vec D \cr }\right),
\end{equation}
where $\vec D=\vec \nabla+ie\vec A(\vec x)$ is the covariant
derivative and $\vec\sigma=(\sigma^1,\sigma^2)$ are the first two
Pauli matrices. This Hamiltonian is obtained in the continuum limit
of some tight-binding lattice models relevant to condensed matter
physics --- on a square lattice with half of a quantum of magnetic
flux through each plaquette (see \cite{semenoff2} and references
therein) or on a honeycomb lattice  without a  magnetic field
\cite{semenoff}.It has also been discussed in the context of the
continuum limit of the {\em d}-wave state of high $T_C$  superconductors
\cite{ssw,mav}.

\subsection{Symmetries}

$H_0$ commutes with any linear combination of the matrices
\begin{equation}
{\cal I}=\left( \matrix{1 & 0 \cr 0 & 1}\right) ,\quad
T^1=\left(\matrix{ 0 & \sigma^3 \cr \sigma^3 & 0 \cr}\right) 
,\quad
T^2=\left(\matrix{ 0 & -i\sigma^3 \cr i\sigma^3 & 0 \cr}\right) 
, \quad 
T^3=\left( \matrix{  1 &   0  \cr  0  &-1 \cr}\right),
\end{equation}
which generate the Lie algebra of a $U(2)$ flavor symmetry which we
will call chiral symmetry. Possible mass terms which can be added to
$H_0$ must be matrices which anti-commute with it. The basic matrix
which anti-commutes with $H_0$ and which  commutes with all of the
flavor generators is 
\begin{equation}
\beta=\left(\matrix{ \sigma^3 & 0 \cr 0 & \sigma^3 \cr }\right)
\end{equation}
and a mass term which one could add to $H_0$ which would preserve
the flavor symmetry is $\beta m$.

Generally, in 2+1 dimensions, fermion mass terms violate parity
symmetry. A parity transformation which is a symmetry of the massless
Dirac  equation  
\begin{equation}
H_0\psi_E(x)=E\psi_E(x)
\end{equation} 
and which commutes with the flavor symmetry generators $T^a$  is
\begin{equation}
\psi_E (x^1,x^2)\rightarrow\left(\matrix{
\sigma^2&0\cr0&-\sigma^2}\right)\psi_E(-x^1,x^2).
\label{pparity}
\end{equation}
The mass term $\beta m$ is odd under this transformation and it
would therefore violate parity symmetry. More generally, all mass
terms of the form $\beta\left( m_0{\cal I}+\vec m\cdot\vec T\right)$ 
are odd under the parity defined in Eq.~(\ref{pparity}). Moreover,
there is no modification of the parity transformation which would
make the mass $\beta m$ parity symmetric.

It is possible to get a parity invariant mass by breaking the flavor
symmetry. For this, we must define parity as a combination of the
transformation (\ref{pparity}) and a discrete flavor transformation,
for example,
\begin{equation}
\psi_E (x^1,x^2)\rightarrow\left(\matrix{
\sigma^2&0\cr0&-\sigma^2}\right)T^1\psi_E(-x^1,x^2).
\label{parity2}
\end{equation}
This is a symmetry of $H_0$ and it also commutes with the mass term
$\beta \left( m_2T^2+m_3T^3\right)$. This mass term breaks the
$U(2)$ flavor symmetry to $U(1)\times U(1)$.

\subsection{Zero modes}

This existence of the unitary matrix $\beta$ which anti-commutes with
$H_0$ implies that $H_0$  has a symmetric spectrum: if $H\psi_E =E
\psi_E$, then $H \beta \psi_E = -E \beta \psi_E$ and, with a
particular choice of phase, $\psi_{-E}=\beta\psi_E$. Thus, for each
positive energy state, there is a negative energy state and the
probability measure  $\psi_E^{\dagger}(x)\psi_E(x)$ is a symmetric
function of $E$.
 
The interesting feature of $H_0$ is the appearance of zero energy
states in a background magnetic field. Consider the equation
\begin{equation}
\vec\sigma\cdot\vec D  u_0=0,
\end{equation}
where the background electromagnetic field is static and has the 
property $\vec \nabla\times \vec A = B(x)$. It has the solution
\begin{equation}
u_0(x)= \exp\left[ ie\phi(x)-e\sigma^3\chi(x) \right] v(x),
\end{equation}
where 
\begin{equation}
-\vec\nabla^2\phi(x)=\vec\nabla\cdot\vec A(x)
,\quad
-\vec\nabla^2 \chi(x)=\vec\nabla\times\vec A(x)=B(x)
\label{flux}
\end{equation} 
and 
\begin{equation}
\vec\sigma\cdot\vec\nabla v(x)=0.
\end{equation}
The second equation in Eqs.~(\ref{flux}) has the solution
\begin{equation}
\chi(x)=-\frac{1}{2\pi}\int d^2y \ln \vert x-y \vert  B(y),
\end{equation}
with asymptotic behavior
\begin{equation}
\lim_{ \vert x \vert \rightarrow\infty}\chi(x)
=-\frac{\chi}{e} \ln \vert x \vert ,
\quad
\chi=\frac{e}{2\pi}\int d^2y B(y).
\end{equation}
(This is strictly correct if the magnetic field goes to zero
sufficiently quickly at infinity for the total flux $\chi$ to be
finite. If it is not finite, there is an infinite number of zero
modes and their density is given by $\chi/V$ where $V$ is the
volume.) Asymptotically, 
\begin{equation}
\lim_{ \vert \vec x \vert \rightarrow\infty} u_0(x)
=\vert\vec x\vert^{\chi\sigma^3} v(x).
\end{equation}
In order to have a normalizable zero mode, the spinor $v(x)$ should
be an eigenvector of $\sigma^3$ with eigenvalue $-\mbox{sgn}(\chi)$
($-1$ if $\chi>0$ and $+1$ if $\chi<0$). It then must be a
normalizable solution of the equations
\begin{equation} 
\sigma^3v(x)=-\mbox{sgn}(\chi)v(x),\qquad
\left( \frac{\partial}{\partial x^1}+
i\mbox{sgn}(\chi)\frac{\partial}{\partial x^2}
\right)v(x)=0.
\end{equation}
Solutions of this equation are $v(x)\sim\left[ x^1
-i\mbox{sgn}(\chi)  x^2\right]^k$ for $k=0,1,\dots,
[\vert\chi\vert]-1$ where $[\vert\chi\vert ]$ is the largest
integer less than $\vert\chi\vert$ and the maximum value of $k$ is
the largest power allowed by  normalizability of the wave function.
Thus there are $[\vert\chi\vert]$ normalizable zero modes. This
fact depends only on the total magnetic flux and  is independent of
the profile of the magnetic field. The result, which we have found
by explicit computation, is also a result of the Atiyah-Singer
index theorem generalized to open spaces\cite{ns}. Some other
applications of this idea are discussed in \cite{semenoff1}.

There are thus $2[\vert\chi\vert]$ zero mode 
wave functions of $H_0$. All of them obey the equation
\begin{equation}
\beta\psi_0=-\mbox{sgn}(\chi)\psi_0,
\label{eig}
\end{equation}
$[\vert\chi\vert]$ of them obey the equation
\begin{equation}
\beta T^3\psi_0=\mbox{sgn}(\chi)\psi_0,
\end{equation}
and $[\vert\chi\vert]$ of them obey
\begin{equation}
\beta T^3\psi_0=-\mbox{sgn}(\chi)\psi_0.
\end{equation}
Of course, any linear combination of them is also a zero mode and
could be chosen to be eigenvalues of any of the flavor symmetry
breaking mass operators $\beta\vec m\cdot\vec T$. All linear
combinations would satisfy Eq.~(\ref{eig}).

\subsection{Interpretation}

The existence of zero modes implies that the Dirac ground state of
the second quantized system of fermions is degenerate. In the ground
state, all negative energy states of the fermions are full and all
positive energy states are empty. For overall charge neutrality,
half of the fermion zero modes must be occupied --- but the
occupation of zero modes is otherwise unconstrained.\footnote{Of
course, occupying half  of the zero modes is only possible when their
number is even, which we shall assume. If it is odd, there are no
neutral ground states.} This means that the degeneracy of the ground
state is $N_{0}!/[(N_{0}/2)!]^2$ where $N_{0}=[\vert\chi\vert]$ is
the number of  zero mode wave-functions. 

In second quantization, the expectation value of any fermion bilinear
operator, using the Dirac commutator for normal ordering, is
\cite{ns}   
\begin{equation}
\left\langle \int d^2x:\Psi^{\dagger}(x)M\Psi(x):\right\rangle
=-\frac{1}{2}\left( \sum_{\rm occupied}\int d^2x
\psi_E^{\dagger}(x)M\psi_E(x)-\sum_{\rm unoccupied}
\int d^2x \psi_E^{\dagger}(x)M\psi_E(x)\right).
\end{equation}
Since, in the Dirac ground state, all positive energy states are
unoccupied  and negative energy states are occupied, and since
$\beta$ maps positive energy states onto negative energy states, if
the matrix $M$ commutes with  $\beta$, only the zero modes survive
in this summation. 

First of all, this means that in a magnetic field the expectation
value of the parity violating mass operator vanishes,
\begin{equation}
\left\langle\int d^2x:\bar\Psi\Psi:\right\rangle
=\left\langle\int d^2x:\Psi^{\dagger}\beta\Psi:\right\rangle=0,
\end{equation}
regardless of the occupation of zero  modes. This has the
implication that external magnetic fields do not enhance the
generation of parity violating mass terms. In fact, it indicates
that if the external field is strong enough that the zero modes are
well isolated, the expectation value of a parity breaking mass
operator, $\langle\int:\Psi^{\dagger}\beta\Psi:\rangle$ should
always vanish.

On the other hand, the parity invariant, chiral symmetry breaking
mass (with $M=\beta T^3$) has the expectation value
\begin{equation}
\left\langle\int d^2x :\bar\Psi(x)T^3\Psi(x):\right\rangle
=-\frac{1}{2}\mbox{sgn}(\chi)\left(
\sum_{\rm occupied}\int d^2x\psi_0^{\dagger}(x)
T^3\psi_0(x)-\sum_{\rm unoccupied}\int d^2x\psi_0^{\dagger}(x)
T^3\psi_0(x)\right).
\end{equation}
It is possible to choose ground states where this expectation value
has  any value between two bounds:
\begin{equation}
-[\vert\chi\vert]
\leq \left\langle\int d^2x :\bar\Psi(x)T^3\Psi(x):\right\rangle
\leq [\vert\chi\vert].
\end{equation}
If we added an infinitesimal mass term $m\beta T^3$ to the
Hamiltonian $H_0$, the system would choose the ground state with
minimal $\beta T^3$, that is the one where\footnote{This is just the
contribution of zero modes. If the total magnetic flux $\chi$ is
finite, the Dirac Hamiltonian has a continuum spectrum whose support
begins at zero energy. Furthermore, there is a threshold density of
continuum states whose contribution to bilinear densities such as
Eq.~(\ref{chrl})  makes up the difference between $[\vert\chi\vert]$
and $\vert\chi\vert$ which is proportional to the spectral
asymmetry of the Hamiltonian and is the correct result. For
details, see \cite{ns}. In the present paper, for simplicity we
ignore the relatively  small error which we make in neglecting the
asymmetry of the continuum spectrum. When the magnetic flux is
infinite, for example in the case of  constant external field,
$[\vert\chi\vert]$ should be replaced by  $\vert eB\vert V/2\pi$
where $V$ is the spatial volume.}
\begin{equation}
\left\langle\int d^2x :\bar\Psi(x)T^3\Psi(x):\right\rangle
=-[\vert\chi\vert]\mbox{sgn}(m\chi) .
\label{chrl}
\end{equation}
The observation that this is so for infinitesimal mass was made in 
\cite{GMS}. However, this does not mean that chiral symmetry is
necessarily broken. It only implies that the ground state is
degenerate and there are chirally non-symmetric  ground states which
are degenerate with chirally symmetric ones. If there are additional
weak interactions, of course, the correct ground state should be
found by resolving the degeneracy using degenerate perturbation
theory. In this case, all that is required to break chiral symmetry 
is an interaction which favors a chirally non-symmetric population 
of the zero modes. 

If the interaction favors the breaking of chiral symmetry,
symmetry breaking occurs for even an infinitesimal value of its
coupling constant. This is the reason why the critical value of the
interaction can be at zero coupling in the presence of a magnetic
field. In fact, in the absence of external magnetic fields a
critical coupling necessary to break chiral symmetry is typically
large. In the presence of the magnetic field it is reduced to
zero. 

For example, it is well known that an interaction of the
NJL model,
\begin{equation}
L_{\rm int}= \frac{\lambda}{2} \left( \bar\Psi T^3\Psi\right)^2,
\end{equation}
when added to the Hamiltonian will break the chiral symmetry if the
coupling constant is attractive and is greater than a particular
critical value. This is seen by analyzing solutions of the
Schwinger-Dyson equation for a mass condensate. It was shown in
\cite{GMS} that, in the presence of a magnetic field, the critical
coupling moves from  some finite value to zero --- even an
infinitesimal coupling will break  the chiral symmetry. That
observation is consistent with our finding here.

An interesting test of this idea would be to examine the effect on
system with the interaction
\begin{equation}
L_{\rm int}=\frac{\kappa}{2}\left( \bar\Psi \Psi\right)^2,
\end{equation}
which can break parity by generating a parity violating fermion mass 
if $\kappa$ has the appropriate sign and is of large enough magnitude
\cite{semenoff2}. Our analysis seems to suggest that a large 
magnetic field should in fact tend to increase the magnitude of the
critical coupling.

\section{Strategy in D=3+1}

In 3+1 dimensions, the situation is more complicated. The zero modes
of the Dirac Hamiltonian in a background magnetic field still have
some dynamics which is non-trivial and there are more possibilities
for  interactions.

To understand the general strategy that we will take, begin with the 
Lagrangian which describes Dirac fermions interacting with an
external electromagnetic field,
\begin{equation}
{\cal S}_0=\int d^4x i\bar\Psi\gamma^\mu D_\mu \Psi ,
\end{equation}
where $D_\mu=\partial_\mu+ieA_\mu(x)$. For concreteness, we choose
the chiral representation of the Dirac matrices,
\begin{equation}
\gamma^0=\left( \matrix{ 0&-i\cr i&0\cr }\right),
\qquad
\gamma^i=i\left(\matrix{ 0&\sigma^i\cr \sigma^i &0\cr}\right),
\end{equation}
and we shall use 
the notation 
\begin{equation}
\Psi=\left( \matrix{ \Psi_L\cr \Psi_R\cr}\right).
\end{equation} 
The Lagrangian has the form
\begin{equation}
{\cal S}_0=\int d^4x \left[ i\Psi_L^{\dagger}\left( D_0+
\vec\sigma\cdot\vec D\right)\Psi_L +i\Psi_R^{\dagger}
\left(D_0-\vec\sigma\cdot\vec D\right)\Psi_R \right].
\label{S_0}
\end{equation}
This action has a global $U(1)_{L}\times U(1)_{R}$ chiral symmetry.
In the following, we must keep in mind that when there are both
background electric and magnetic fields, or if the  gauge field is
dynamical, this symmetry is reduced to the vector $U(1)$ by the
axial anomaly. For simplicity, in this paper we will assume that
this is not the case. It would be straightforward to extend the
present work to models with dynamical $U(1)$ fields, by introducing
more species of fermions so that there exist chiral symmetries which
are unaffected by the axial anomaly. For now, we will consider the
case of a constant non-dynamical  background magnetic field,
\begin{equation}
A_0=0,\quad
A_i= -\frac{B}{2}\epsilon_{ij3}x^j .
\label{dmsu1}
\end{equation}
In this background field, it is convenient to make use of the
mixed spacetime-momentum representation as follows:
\begin{equation}
{\cal S}_0=\int \frac{d\omega dk}{(2\pi)^2} d^2z\left[
\Psi_{L}^{\dagger}\left(\omega -k\sigma^3+{\cal D}\right)\Psi_{L} 
+i\Psi_{R}^{\dagger}\left(\omega + k\sigma^3-{\cal D}\right)\Psi_{R}
\right], 
\end{equation}
where, by definition, $z=x^1+ix^2$, $d^2z=dx^1 dx^2$ and 
\begin{equation}
{\cal D}= i \left( 
\matrix{ 0& 2\frac{\partial}{\partial z}+\frac{eB}{2} \bar{z}
\cr 2\frac{\partial}{\partial \bar{z}}-\frac{eB}{2} z&0 
\cr }\right).
\end{equation}
The spectrum of $(k\sigma^3-{\cal D})$ is well known. The equation
\begin{equation}
\left( k\sigma^3-{\cal D}\right)\phi_\lambda(\bar{z},z)
=\lambda\phi_\lambda (\bar{z},z)
\end{equation}
has eigenvalues 
\begin{equation}
\lambda =\pm\sqrt{k^2+2n|eB|},
\end{equation}
for all integers $n=0,1,2,\dots$ . When $n\geq 1$, these are
dispersion relations of (1+1)-dimensional Dirac fermions with masses
given by $\sqrt{2n\vert eB\vert}$. [Of course, this mass gap is not
Lorentz  invariant from a four-dimensional point of view. However, it
is Lorentz invariant in  1+1 dimensions. The reason for this is that
there is a subgroup of the Lorentz group which survives in the
background magnetic field --- it is invariant under boosts in the
direction of the field lines. Here this will mean  that the 
effective theory for Landau levels must be invariant under the
(1+1)-dimensional Lorentz transformations.]  

Zero modes of ${\cal D}$ are  solutions of the equation
\begin{equation}
{\cal D}\phi_{m}(z,\bar{z})=0,
\end{equation}
and are given by the infinite set of ortho-normal functions
\begin{equation}
\phi_{m}(z,\bar{z})= \frac{\bar{z}^{m}}{\sqrt{\pi\Gamma(m+1)}}
\left( \frac{eB}{2}\right)^{(m+1)/2} 
\exp\left(-\frac{eB}{4}z\bar{z}\right) 
\left( \matrix{ 0 \cr 1 \cr} \right),
\end{equation}
with $m=0,1,2,\dots$ . Here, without loss of generality, we assumed 
that $eB>0$. In the (1+1)-dimensional theory, the zero modes of 
${\cal D}$ correspond to massless fermions. The Lagrangian is
(restoring the space dependence) 
\begin{equation}
{\cal S}_0=\int dx^0 dx^3 \sum_{m=0}^\infty \left( 
i\psi_m^{(L)*} \partial_{+}\psi_m^{(L)} 
+ i\psi_m^{(R)*}\partial_{-}\phi_m^{(R)}\right)
+(\mbox{massive modes}),
\label{massless}
\end{equation}
where $\partial_\pm\equiv \partial_0\mp\partial_3$ and the new
fields are defined as the coefficient functions in the expansions
of $\Psi_{L}$ and $\Psi_{R}$ over the complete set of eigenstates:
\begin{eqnarray}
\Psi_{L}&=&\sum_{m=0}^{\infty}\psi_m^{(L)}(x^0,x^3)
\phi_{m}(z,\bar{z}) +\sum (\mbox{massive modes}), 
\label{psi-psi-L}\\
\Psi_{R}&=&\sum_{m=0}^{\infty}\psi_m^{(R)}(x^0,x^3)
\phi_{m}(z,\bar{z})+\sum (\mbox{massive modes}).
\label{psi-psi-R}
\end{eqnarray}

Note that the kinetic term in the action (\ref{massless}) for the
massless modes appears to have a $U(N)_{R}\times U(N)_{L}$ with
$N\rightarrow\infty$ symmetry. This is just the unitary symmetry
which mixes the different wave functions of the degenerate
zero modes, and which is actually there for every Landau level.
This effective symmetry is not preserved by interactions.

We will analyze the possibility that interactions that are added to
this field theory drive a spontaneous breaking of the $U(1)_{R}\times
U(1)_{L}$ chiral symmetry by generating a mass gap for the fermions
in Eq.~(\ref{massless}). We expect that this spontaneous symmetry
breaking takes place at very long wavelengths. Our approach to
the problem in the following sections can be summarized as follows:
\begin{enumerate}

\item{}We consider the theory of four-dimensional fermions in a 
magnetic field as described above, with some interactions which 
should be local and respect (3+1)-dimensional Poincar\'{e} and
chiral symmetry but are otherwise unspecified.

\item{}We then consider the effective field theory which is obtained
by integrating out all of the massive modes of the fermions in
Eq.~(\ref{massless})  and momentum states of the massless modes
above an ultraviolet cutoff. 

\item{}We assume that the resulting effective Lagrangian is local. 
To guarantee locality, we generally have to  assume that the
interactions between fermions in four dimensions are
short ranged.\footnote{If there are long-ranged interactions which
are mediated by massless fields, the correct procedure would be to
retain the long wavelength modes of the massless fields in the
effective Lagrangian.}   Furthermore, the ultraviolet cutoff for the
effective theory should be less than the mass gap of the lightest
massive mode, $\vert e B\vert$. 

\item{}We consider all relevant operators which could be added to
the effective Lagrangian which are consistent with the symmetries of
the theory. Since they should be relevant in the sense of two
dimensional field theory, these are all possible four-fermion
operators which are consistent with symmetry. There are an infinite
number of such operators.

\item{}We compute the beta function for the coupling constants of the 
relevant operators and look for infrared stable fixed points. These 
fixed points should govern the behavior of the very long-wavelength
degrees of freedom of the theory.

\item{}If the coupling constants flow to an infrared stable fixed
point, then an infrared limit of the massless theory exists and there
is no symmetry breaking. If, on the other hand, the coupling
constant flow is not in the domain of attraction of any infrared
fixed point, so that it flows to strong coupling in the infrared, we
postulate that this implies the dynamical generation of a mass
gap --- and  spontaneously broken chiral symmetry. We shall find
examples of both kinds of behavior.

\end{enumerate}

An example of a local four dimensional interaction which preserves
the $U(1)_{R} \times U(1)_{L}$ chiral symmetry is the
NJL interaction \cite{NJL},
\begin{equation}
{\cal S}_{int}=\frac{G}{2}\int d^4x \left[(\bar{\Psi}\Psi)^2
+(\bar{\Psi}i\gamma_5\Psi)^2\right].
\label{lagNJL}
\end{equation}
A renormalizable version of this interaction would be one which is 
mediated by a massive scalar mesons.

\section{General structure of the low-energy theory}

The constraints of (1+1)-dimensional Lorentz invariance and
$U(1)_{R}\times U(1)_{L}$ chiral symmetry allow four-fermion 
coupling constants as in the effective theory,
\begin{equation}
{\cal L}_{eff} =\sum_{n=0}^{\infty}\left(
\psi^{(L)*}_{n} i \partial_{+}\psi^{(L)}_{n}
+\psi^{(R)*}_{n} i \partial_{-}\psi^{(R)}_{n}\right)
+\sum_{n_1 , n_2, m_1, m_2=0}^{\infty}
g_{0}\left(\begin{array}{cc}
             n_1 & n_2 \\
             m_1 & m_2   
       \end{array}\right)
\psi^{(L)*}_{n_1} \psi^{(R)}_{n_2}
\psi^{(R)*}_{m_1} \psi^{(L)}_{m_2}.
\label{eff-gen}
\end{equation}
The coupling constants $g_{0}\left(\begin{array}{cc}
             n_1 & n_2 \\
             m_1 & m_2   
       \end{array}\right)$ 
obey further constraints from charge conjugation, parity and time
reversal (CPT) symmetry and the symmetry  of the underlying theory
under translations and rotations about the axis of the magnetic
field which are summarized in Appendix~\ref{AppA}. A general
solution of those constraints would yield the most general allowed
structure of the low-energy effective action in
Eq.~(\ref{eff-gen}). 

Some particular solutions of those constraints are of interest.
For example, the constraints allow the maximally symmetric solution
with  $U(N)_{R}\times U(N)_{L}$ ($N\to\infty$) symmetry, 
\begin{equation}
g\left(\begin{array}{cc} 
             n_1 & n_2 \\
             m_1 & m_2   
       \end{array}\right)
=g_{LR}\delta_{n_1,m_2}\delta_{n_2,m_1},
\label{UnUn}
\end{equation}
where $g_{LR}$ is real. Similarly, there is a solution with 
$U(N)_{V}$ (with $N\to\infty$) symmetry, 
\begin{equation}
g\left(\begin{array}{cc}
             n_1 & n_2 \\
             m_1 & m_2
       \end{array}\right)
=g_{V}\delta_{n_1,n_2}\delta_{m_1,m_2},
\label{Uv}
\end{equation}
where $g_{V}$ is real again.

In addition a large class of solutions could be found by the
reduction of the interactions in the original (3+1)-dimensional 
model to the lowest Landau level. After such a reduction of the
NJL interaction (\ref{lagNJL}), we arrive at 
\begin{equation}
g\left(\begin{array}{cc}
             n_1 & n_2 \\
             m_1 & m_2
       \end{array}\right)
=\frac{G}{2^{(n_1+m_1+n_2+m_2)/2}}
\sqrt{\frac{ \Gamma(n_1+m_1+1)\Gamma(n_2+m_2+1)}
{\Gamma(n_1+1)\Gamma(m_1+1)\Gamma(n_2+1)\Gamma(m_2+1)}}.
\label{lll-int}
\end{equation}
Since it descends from a Lorentz and chirally invariant interaction
in 3+1 dimensions, it must necessarily satisfy the constraints of
symmetry [see Eqs.~(\ref{real}), (\ref{parity}), (\ref{restrict1}),
(\ref{restrict2}) and (\ref{nnmm})]. Unlike the first two solutions
in Eqs.~(\ref{UnUn}) and (\ref{Uv}), this last one in
Eq.~(\ref{lll-int}) does not seem to have any extra symmetry in
addition to the required $U(1)_{R}\times U(1)_{L}$ flavor symmetry.

Now, we envisage having obtained the effective action
(\ref{eff-gen}) by integrating out all modes in the higher Landau
levels, as well as all momentum modes of the fermions in the lowest
Landau level which are above a certain cutoff. Of the many
interactions that this procedure would produce, we have kept only
the local four-fermion operators. This procedure is  legitimate
only if the ultraviolet cutoff of this model is lower than  the
lowest mass gap of the fields which have been eliminated, i.e.
$\vert eB\vert$. By chiral symmetry and Lorentz invariance, the
effective action cannot contain mass terms for the fermions.
Furthermore, the only Lorentz invariant four-fermion operator is of
the form given  in Eq.~(\ref{eff-gen}).

The renormalization group procedure examines how the coupling
constants in Eq.~(\ref{eff-gen}) change as we further lower the
cutoff to isolate the very long wavelength excitations. This
information is encoded in the  beta function.

The $\beta$ function for the general coupling constant in
Eq.~(\ref{eff-gen})  is computed in $2+\epsilon $ dimensions to
two loop  order in Appendix~\ref{AppB}. The result is 
\begin{eqnarray} 
\beta(g)\left(\begin{array}{cc} 
             N_1 & N_2 \\
             M_1 & M_2   
       \end{array}\right)&&
= 2 \varepsilon g\left(\begin{array}{cc} 
             N_1 & N_2 \\
             M_1 & M_2  
        \end{array}\right)
-\frac{1}{2\pi}\sum_{k_1}\left[
g\left(\begin{array}{cc}
             N_1 & N_2 \\
             k_1 & N_1-N_2+k_1     
       \end{array}\right)
g\left(\begin{array}{cc}
             N_1-N_2+k_1 & k_1 \\
             M_1 & M_2     
       \end{array}\right)\right.
\label{beta2loop} \\
&&-\left.g\left(\begin{array}{cc}
             N_1 & k_1 \\
             M_1 & N_1+M_1-k_1
       \end{array}\right)
g\left(\begin{array}{cc} 
             N_1+M_1-k_1 & N_2 \\
             k_1 & M_2     
       \end{array}\right)\right]
\nonumber\\[3mm]
-\frac{1}{8\pi^2}
\sum_{k_1,k_2}&&  \left[
g\left(\begin{array}{cc}
             N_1 & k_2 \\
             k_1 & N_1+k_1-k_2    
       \end{array}\right)
g\left(\begin{array}{cc}
             N_1+k_1-k_2 & N_2 \\
             M_1& M_2+k_1-k_2     
       \end{array}\right)
g\left(\begin{array}{cc}
             M_2+k_1-k_2 & k_1 \\
             k_2 & M_2    
       \end{array}\right)\right.
\nonumber\\
&&+\left.g\left(\begin{array}{cc}
             k_1 & k_2 \\
             M_1 & M_1+k_1-k_2     
       \end{array}\right)
g\left(\begin{array}{cc}
             N_1 & N_2-M_1+k_2 \\
             k_2 & M_2    
       \end{array}\right)
g\left(\begin{array}{cc}
             M_1+k_1-k_2 & N_2 \\
             N_2-M_1+k_2 & k_1    
       \end{array}\right) \right]
\nonumber\\[3mm]
+\frac{2}{(4\pi)^2}
 g\left(\begin{array}{cc}
             N_1 & N_2 \\
             M_1 & M_2     
       \end{array}\right)
&&\sum_{k_1,k_2} \left[
g\left(\begin{array}{cc}
             N_1 & k_2+N_1 \\
             k_1 & k_1-k_2    
       \end{array}\right)
g\left(\begin{array}{cc}
             k_1-k_2  & k_1 \\
             k_2+N_1 & N_1    
       \end{array}\right)
+g\left(\begin{array}{cc}
             k_1 & k_2+N_2 \\
             N_2 & k_1-k_2     
       \end{array}\right)
g\left(\begin{array}{cc}
             k_1-k_2 & N_2 \\
             k_2+N_2 & k_1     
       \end{array}\right)\right.\nonumber\\
&&+\left.g\left(\begin{array}{cc}
             k_1 & k_2+M_1 \\
             M_1 & k_1-k_2     
       \end{array}\right)
g\left(\begin{array}{cc}
             k_1-k_2 & M_1 \\
             k_2+M_1 & k_1     
       \end{array}\right)
+g\left(\begin{array}{cc}
             M_2 & k_2+M_2 \\
             k_1 & k_1-k_2    
       \end{array}\right)
g\left(\begin{array}{cc}
             k_1-k_2 & k_1 \\
             k_2+M_2 & M_2    
       \end{array}\right)\right].
\nonumber
\end{eqnarray}
 
Finally, before we proceed to the next section, we note that we may
have to occasionally cut off the summation over modes in the first
Landau level. We do this by summing modes up to some maximum number,
$n=0,1,\dots,N$. Such a situation could be produced by considering
the fermions in an external  magnetic field with finite magnetic
flux, or with uniform field and finite  transverse area. In that
case, $N$ which can be thought of as  the number of ``flavors" which
goes to infinity as  $|eB|S_{12}/2\pi$ where $S_{12}\to\infty$ is the
area of the two-dimensional perpendicular (with respect to the
direction of the magnetic field) subspace. 

\section{Symmetric fixed points}

In Appendix~\ref{AppB}, we have computed the two-loop
$\beta$ function  (\ref{beta2loop}) for the most general coupling
constants, subject to the restriction in Eq.~(\ref{nnmm}). The next
natural step would be locate infrared stable fixed points of this
beta function. This problem is far more complicated than we can
solve at present. Instead, in our analysis below, we shall restrict
our attention to the beta function for coupling constants with close
to maximal symmetry.

For some scalar field theories, where there are many components of a
scalar  field and renormalizable interactions which couple them, it
is known that the only infrared fixed points of the renormalization
group flow of relevant couplings are those with maximal symmetry
\cite{BGZ,ZJ}. To our knowledge, there is no similar theorem for
Gross-Neveu like models. On the other hand, we consider it plausible
that similar arguments can be applied: in particular, if no maximally
symmetric infrared stable fixed point exists, then there are no
infrared stable fixed points at all. 

Maximally symmetric couplings can easily be shown to be contained in
other combinations of coupling constants. In Appendix~\ref{AppC} we
show that the NJL coupling contains the maximally symmetric ones.

Consider the renormalization group flow in the special case when
$U(N)_{R}\times U(N)_{L}$ and $U(N)_{V}$ couplings, as defined in
Eqs.~(\ref{UnUn}) and (\ref{Uv}), are the only non-zero couplings in
the effective action. Then, from our general result in the previous
section, we extract the following expressions for the $\beta$
functions of interest:
\begin{eqnarray}
\beta_{LR}&=&2\varepsilon g_{LR}+\frac{1}{2\pi} g_{V}^2
+\frac{1}{4\pi^2} \left(N g_{LR}^3 +2 g_{V} g_{LR}^2  
+N g_{V}^2 g_{LR} -g_{V}^3\right), \label{beta_LR}\\
\beta_{V}&=&2\varepsilon g_{V}-\frac{N}{2\pi} g_{V}^2
+\frac{1}{4\pi^2} \left[ 2N g_{V}^3 -(N-3) g_{V}^2 g_{LR} 
-(N-2) g_{V} g_{LR}^2 \right]. \label{beta_V}
\end{eqnarray}
Note that we have been forced to introduce a cutoff on $N$. In order
to get a sensible result when taking the limit $N\to\infty$, it is
convenient to rescale the couplings as follows: $g_{LR} \to g_{lr}
/\sqrt{N}$ and $g_{V} \to g_{v}/N$. As is clear from the definition,
the $\beta$ functions will also get rescaled accordingly,
$\beta_{LR} \to \beta_{lr}/\sqrt{N}$ and $\beta_{V}  \to
\beta_{v}/N$. After taking this into account and performing the
limit $N\to\infty$, we arrive at
\begin{eqnarray}
\beta_{lr}&=&2\varepsilon g_{lr} +\frac{1}{4\pi^2}  g_{lr}^3 , 
\label{beta_lr}\\
\beta_{v}&=&2\varepsilon g_{v}-\frac{1}{2\pi} g_{v}^2
+\frac{1}{4\pi^2} g_{v} g_{lr}^2 . \label{beta_v}
\end{eqnarray}
It must be emphasized that we have adopted the rescaling of the
most symmetric $g_{LR}$ coupling by $1/\sqrt{N}$ rather than $1/N$. 
Since the rescaling performed above led to the well-defined
$\beta$ functions, we conclude that the (1+1)-dimensional action in
Eq.~(\ref{eff-gen}) with $g_{LR}$ of order $\sim 1/\sqrt{N}$
describes a consistent and non-trivial interacting theory in the
limit $N\to\infty$. \footnote{As one can see, the rescaling of
$g_{LR}$ by $1/N$ is also meaningful. The resulting
$\beta$ functions are
\begin{eqnarray}
\beta_{lr}&=&2\varepsilon g_{lr} , 
\label{b_lr}\\
\beta_{v}&=&2\varepsilon g_{v}-\frac{1}{2\pi} g_{v}^2.
\label{b_v}
\end{eqnarray}
The corresponding theory is less interesting. Indeed, the
expressions for the $\beta$ functions in Eq.~(\ref{b_lr}) and
(\ref{b_v}) in the limit $\varepsilon =0$ describe the situation
when the $g_{lr}$ coupling does not run at all, while the $g_{v}$
experiences asymptotic freedom from the $g_{v}>0$ side and 
infrared freedom from the $g_{v}<0$ side. We shall see in a moment
that this picture corresponds to a special case ($g_{lr}=0$) of the
flow described by Eqs.~(\ref{beta_lr}) and (\ref{beta_v}).}

\begin{figure}
\epsfbox{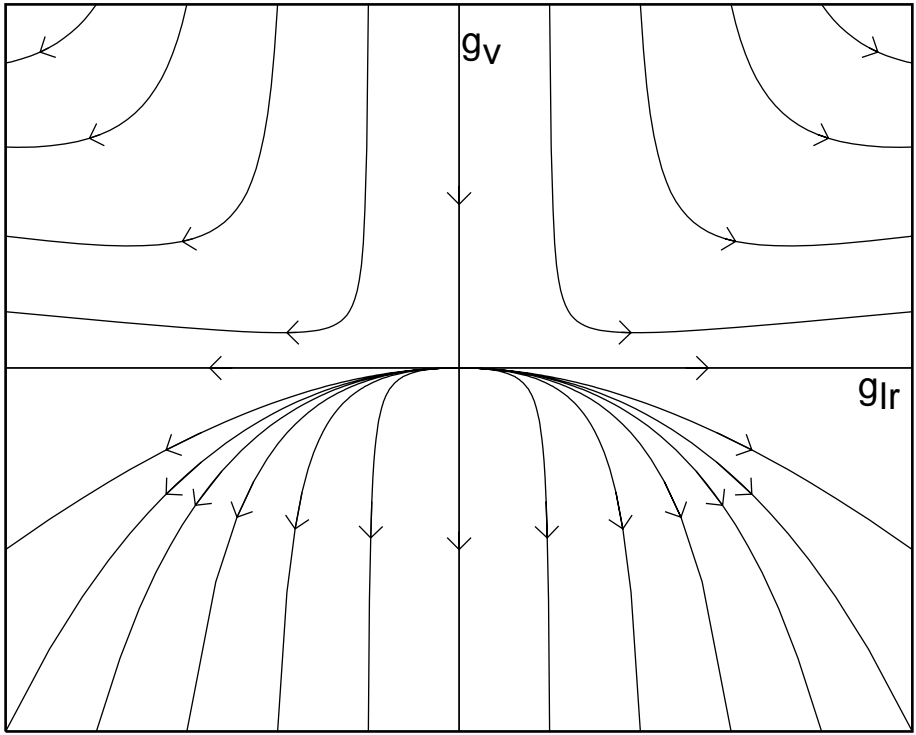}
\caption{The renormalization group flow in the $(g_{lr},g_{v})$ plane.}
\label{fig6}
\end{figure}

When $\varepsilon=0$ we can solve the renormalization group
equations (\ref{beta_lr}) and (\ref{beta_v}) explicitly, and the
analytical solution reads
\begin{eqnarray}
g_{lr}(t)&=&\frac{g_{lr}(0)}{\sqrt{1-\frac{g^2_{lr}(0) t}
{2\pi^2}}}, \label{sol-lr}\\
g_{v}(t)&=&\frac{g_{v}(0)}{\left(1+\frac{2\pi g_{v}(0)}
{g^2_{lr}(0)}\right) \sqrt{1-\frac{g^2_{lr}(0) t}{2\pi^2}}
-\frac{2\pi g_{v}(0)}{g^2_{lr}(0)}
\left(1-\frac{g^2_{lr}(0) t}{2\pi^2}\right)} 
\label{sol-v}.
\end{eqnarray}
This flow is presented graphically in Fig.~\ref{fig6} where the
arrows show the flow direction toward ultraviolet. 

In the upper half-plane $g_{v}>0$, the simple analysis of the flow
given by Eqs.~(\ref{sol-lr}) and (\ref{sol-v}) reveals infrared
(with $g_{v}\to+\infty$) and ultraviolet (with $g_{lr}\to\pm\infty$
and $g_{v}\to+\infty$) Landau poles at  
\begin{eqnarray}
t_{IR}&=&-\left(
\frac{2\pi}{g_{v}(0)}+\frac{g^2_{lr}(0)}{2g_{v}^2(0)} 
\right),\\
t_{UV}&=&\frac{2\pi^2}{g^2_{lr}(0)} ,
\end{eqnarray}
respectively. We argue that the strong infrared dynamics (with
$g_{v}\to+\infty$) in this half-plane of couplings is an
indication of a mass generation and breaking of the chiral $U(1)$
symmetry. Indeed, the generation of the fermion mass in the
infrared region seems to be the only way one can avoid running into
the problem of the physical Landau pole. 

The generation of the fermion mass in the $g_{v}>0$ half-plane, in
its turn, is consistent with the expectation of the universality of
the magnetic catalysis in a wide range of (3+1)-dimensional models
(such as the NJL) with a short range interaction. Indeed, as we
established, the low energy dynamics in such models is described by
the (1+1)-dimensional effective action in Eq.~(\ref{eff-gen}) with
the coupling satisfying the set of restriction in
Eqs.~(\ref{restrict1}), (\ref{restrict2}) and (\ref{nnmm}). The
generic coupling [say, like that in Eq.~(\ref{lll-int}) coming from
the interaction of the lowest Landau level modes] which does not
have any extra symmetry is still expected to have the
$U(N)_{R}\times U(N)_{L}$ and $U(N)_{V}$ contribution [see
Eqs.~(\ref{gLLL-LR}) and (\ref{gLLL-V})]. Then, if this $U(N)_{V}$
contribution is positive, $g_{v}>0$, it is going to drive the
system to the generation of mass. 

Now let us study the flow in the lower half-plane $g_{v}\leq 0$.
As is easy to see, there is an infrared fixed point at
$(g_{lr},g_{v})=(0,0)$ and the ultraviolet Landau pole  (with
$g_{v}\to-\infty$ while $g_{lr}$ is either fixed or  approaches
$\pm\infty$) at the following values of $t$:
\begin{eqnarray}
t_{UV}&=&\left( \frac{2\pi}{\vert g_{v}(0)\vert}
-\frac{g^2_{lr}(0)}{2g_{v}^2(0)} \right), 
        \quad \mbox{if} \quad 
\vert g_{v}(0)\vert\geq\frac{g^2_{lr}(0)}{2\pi},\\
t_{UV}&=&\frac{2\pi^2}{g^2_{lr}(0)} ,
        \quad \mbox{if} \quad 
\vert g_{v}(0)\vert < \frac{g^2_{lr}(0)}{2\pi}.
\end{eqnarray}
Since the infrared fixed point $(0,0)$ corresponds to weakly
coupled dynamics, there is apparently no mass generation in this
half-plane of the coupling space. This is also in full agreement
with our general expectation. Indeed, the negative values of
$g_{v}$ correspond to the repulsion rather than attraction in the
fermion-antifermion channel which is responsible for the generation
of mass, breaking chiral $U(1)$ symmetry.

Our conclusion here is that whether this model breaks chiral
symmetry or not is entirely dependent on the sign of the coupling
constant $g_v$.

The reader might be puzzled by the fact that a $U(1)$ chiral
symmetry can be broken in an effectively two dimensional system.
We emphasize here that this phenomenon is identical to that in the
chiral Gross-Neveu model. Strictly speaking, chiral symmetry can
only be broken  in the large $N$ limit. The finite $N$  system
should still be chirally symmetric, as the low dimensionality of
the system would not allow for spontaneous breaking of a continuous
symmetry. Indeed, if we consider the case of a large but finite
value of $N$, an infrared stable fixed point appears for $g_v>0$ at
$(g_{lr},g_{v}) \simeq (-2\pi,\pi N)$ which goes to infinity with
$N\to\infty$. This fixed point is at strong coupling, so a
conclusion based on perturbation theory is speculative at best, but
its appearance  is consistent with the expectation that  the
massless limit of this model is well defined (albeit strongly
coupled) when $g_v>0$ and $N$ is finite and chiral symmetry breaking
need not take place. When $N\rightarrow\infty$ this fixed point
moves to infinite coupling and chiral symmetry breaking is possible.

\section{Conclusion}

In this paper we have shown that the effective low-energy dynamics
of the $U(1)$ chirally symmetric models with a short range
interaction in a background magnetic field is described by a
(1+1)-dimensional Gross-Neveu like model with an infinite number of
flavors [see Eq.~(\ref{eff-gen})]. Here we established that different
flavors come out as the representation space of the magnetic
translations in the original (3+1)-dimensional model. The number of
flavors is infinite and proportional to the area of the two
dimensional space perpendicular to the magnetic field. 

Based exclusively on the arguments of symmetry, we established a set
of conditions, given by Eqs.~(\ref{real}), (\ref{parity}),
(\ref{restrict1}), (\ref{restrict2}) and (\ref{nnmm}), that the
couplings of the effective theory have to satisfy. To show that they
allow a non-trivial solution, we presented a few (out of infinitely
many possible) examples of couplings that satisfy all the constrains.
Among them, there are, in particular, the highly symmetric
$U(N)_{R}\times U(N)_{L}$ and $U(N)_{V}$ (with $N\to\infty$)
couplings. These latter are of special interest because their
renormalization group flow is self-contained and allows an analytical
solution [see Eqs.~(\ref{sol-lr}) and (\ref{sol-v})]. 

At the level of the effective theory, we calculated the two-loop
$\beta$ function and analyzed the renormalization group flow in  the
two-dimensional subspace of the $U(N)_{R}\times U(N)_{L}$ and
$U(N)_{V}$ couplings. The general result is argued to indicate
the generation of the fermion mass in the $g_{v}>0$ half-plane
of couplings. In the other half-plane, the infrared dynamics is
weakly coupled and there is no mass generation. This mass generation
pattern is consistent with the earlier suggested conjecture of the
universality of the so-called magnetic catalysis \cite{GMS}
in, at least, the models from the same universality class as the
chiral $U(1)$ NJL model.

\begin{acknowledgments}
We would like to thank V.~Miransky for many interesting discussions.
The work of G.W.S. was supported by the Natural Sciences and
Engineering Research Council of Canada and by the Niels Bohr Fund of
Denmark. The work of I.A.S. and L.C.R.W. was supported  by the U.S.
Department of Energy Grant No. DE-FG02-84ER40153.
\end{acknowledgments}

\appendix

\section{Constraints on coupling constants}
\label{AppA}

Let us clarify the origin of the action as well as the meaning of
different flavors in Eq.~(\ref{eff-gen}) in terms of the original
(3+1)-dimensional model defined in Eqs.~(\ref{S_0}) and
(\ref{lagNJL}). 

We start from the analysis of the space-time symmetries in the
model. Notice that, due to the presence of the background field, the
standard translations in the two-dimensional plane perpendicular
to the magnetic field are not symmetries of the original model in
Eq.~(\ref{S_0}). Nevertheless, there are other transformations, the
so-called magnetic translations, which leave the action invariant. In
contrast to the case of the ordinary translations, the two generators
of the magnetic translations do not commute. In the symmetric gauge
given in Eq.~(\ref{dmsu1}), the explicit representation of the
generators and their commutation relation read    
\begin{mathletters}
\begin{eqnarray}
X&=&\sqrt{eB}\left(x^{1}+i\frac{D_2}{eB}\right), 
\label{X&Ya}\\
Y&=&\sqrt{eB}\left(x^{2}-i\frac{D_1}{eB}\right), 
\label{X&Yb}\\ 
&&\left[X,Y\right]=i, \label{X&Yc} 
\end{eqnarray}
\end{mathletters}
where we assume that $eB>0$. It is easy to check that the operators
$X$ and $Y$ commute with the Hamiltonian of our model. Instead of
these $X$ and $Y$, it is  convenient to introduce the ``creation"
and ``annihilation" operators
\begin{mathletters}
\begin{eqnarray}
a&=&\frac{X+iY}{\sqrt{2}}, \label{a&a}\\
a^{\dagger}&=&\frac{X-iY}{\sqrt{2}},\label{a&b}\\ 
&&\left[a,a^{\dagger}\right]=1. \label{a&c}
\end{eqnarray}
\end{mathletters}
Having introduced these operators in the problem, we realize
that the Fock space is spanned by the set of states
$|q,m\rangle$ where the quantum number $m=0,1,2, \dots$ denotes
the eigenvalue of $a^{\dagger}a$ operator and the multi-index
$q$ represents all the other  quantum numbers (say, the Landau
level number, the fermion spin projection  and the chirality). 
In the absence of any vacuum rearrangement (symmetry breaking), 
the above set of states (in coordinate representation) reads 
\begin{equation} 
\langle x | m,n,\sigma,\chi,p_{\parallel} \rangle =
\frac{1}{\sqrt{2\pi}l}\frac{1}{(\sqrt{2}l)^{m-n}}
\sqrt{\frac{n!}{m!}} e^{-ix_{\parallel} p_{\parallel}} 
\bar{z}^{m-n} L^{(m-n)}_{n}\left(\frac{z\bar{z}}{2l^2}\right)
\exp\left(-\frac{z\bar{z}}{4l^2}\right)\phi_{\sigma,\chi},
\label{states}
\end{equation}
where $l=1/\sqrt{eB}$ is the magnetic length, $z=x^{1}+ix^{2}$,
$\bar{z}=x^{1}-ix^{2}$, $x_{\parallel} p_{\parallel}=x^0 p_0-x^3p_3$
and $\phi_{\sigma,\chi}$ is  the spinor with a given spin projection
$\sigma$ and chirality $\chi$. Note that the expression in
Eq.~(\ref{states}) is well defined even for the case  when $m<n$
(both numbers are positive) due to the Rodrigues formula for the
generalized Laguerre polynomials, 
\begin{equation}
z^{m-n}L^{(m-n)}_{n}(z) = \frac{1}{n!}e^{z}\frac{d^n}{d z^n}
\left(e^{-z} z^m\right).
\end{equation}
One has to remember also that all the modes in the lowest Landau
level ($n=0$) have the same projection of the spin, while the modes
in the higher Landau levels ($n\geq 1$) have both projections. 

Since we are interested in the structure of the effective action,
describing the infrared dynamics ($p_0 \ll \sqrt{eB}$), it is
sufficient to take into account only those degrees of freedom that
originate from the lowest Landau level modes ($n=0$). These modes
freely propagate in the (1+1)-dimensional parallel  $(x^{0},x^{3})$
subspace, and  are classified by the chirality and the eigenvalue of
$a^{\dagger}a$ [see Eq.~(\ref{states})]. In what follows, we denote
the effective degrees of freedom accordingly, $\psi^{(L,R)}_{m}
(x_{\parallel})$, where the superscript $(L)$ or $(R)$ denotes the
states that result from the (3+1)-dimensional states of definite
chirality. In the parallel subspace $(x^{0},x^{3})$, $\psi^{(L)}_{m}
(x_{\parallel})$ and $\psi^{(R)}_{m} (x_{\parallel})$ have the
interpretation of the left and right moving along the $x^{3}$ axis
modes, respectively.

While restricting the kinetic term in the NJL model (\ref{lagNJL})
to the lowest Landau level modes, 
\begin{equation}
\Psi\left(x_{\parallel},x_{\perp}\right)\to
\sum_{m=0}^{\infty}\frac{\phi_{+,+}
\psi^{(R)}_{m} (x_{\parallel})+\phi_{+,-}
\psi^{(L)}_{m} (x_{\parallel}) }{\sqrt{2\pi m!} l} 
\left(\frac{\bar{z}}{\sqrt{2}l}\right)^{m}
\exp\left(-\frac{z\bar{z}}{4l^2}\right) ,
\label{psi-LR}
\end{equation}
we check that $\gamma^{\mu}\Pi_{\mu} \to \gamma_{\parallel}^{\mu}
p^{\parallel}_{\mu}$. After performing the integration over the
perpendicular space coordinates in the original NJL action, we
arrive at the effective model as in Eq.~(\ref{eff-gen}). Remarkably,
while the NJL coupling $G$ is dimensionful, the effective coupling
in Eq.~(\ref{eff-gen}) is dimensionless, $g\sim G/l^2\equiv G|eB|$. 

From the derivation above, we see that the fields of different
flavors in the effective (1+1)-dimensional theory (\ref{eff-gen})
correspond to different eigenstates of $a^{\dagger}a$ operator. This
simple observation, as we show below, has far reaching consequences.

Now, let us establish the allowed structure of coupling constants in
Eq.~(\ref{eff-gen}). The most general restriction on the couplings
comes from the condition of reality of the action. This requires
that   
\begin{equation}
g^{*}\left(\begin{array}{cc}
             n_1 & n_2 \\
             m_1 & m_2   
       \end{array}\right)
=g\left(\begin{array}{cc}
             m_2  & m_1 \\
             n_2 & n_1  
       \end{array}\right).
\label{real}
\end{equation}
Similarly, the invariance under the parity ($x^3\to -x^3$) leads to
another restriction: 
\begin{equation}
g\left(\begin{array}{cc}
             n_1 & n_2 \\
             m_1 & m_2   
       \end{array}\right)
=g\left(\begin{array}{cc}
             m_1 & m_2  \\
             n_1 & n_2  
       \end{array}\right).
\label{parity}
\end{equation}
These two conditions are too general and so are not of great
interest or of great power by themselves. It turns out, however,
that there are other, more restrictive conditions. 

By recalling the origin of flavors in Eq.~(\ref{eff-gen}), we
realize that the effective theory should enjoy some kind of flavor
symmetry that results from the symmetry under magnetic translations
of the original (3+1)-dimensional theory. This flavor symmetry, as
will become clear in a moment, puts further restrictions on the
allowed structure of the four-index coupling in Eq.~(\ref{eff-gen}). 

The infinitesimal transformations of the magnetic translations
and the related rotation in the perpendicular plane are given by the
following operators:
\begin{mathletters}
\begin{eqnarray}
U_{1}&=&1+i\varepsilon_{1} X \equiv 
1 +\frac{i\varepsilon_{1}}{\sqrt{2}}\left(a+a^{\dagger}\right),
\label{trans1}\\
U_{2}&=&1+i\varepsilon_{2} Y \equiv
1 +\frac{\varepsilon_{2}}{\sqrt{2}}\left(a-a^{\dagger}\right), 
\label{trans2}\\
U_{12}&=&1+i\varepsilon_{12} a^{\dagger}a.
\label{trans12}
\end{eqnarray}
\end{mathletters}
To determine the transformation properties of the fields of
different flavors, we again recall that, by construction, these
fields are the eigenstates of the $a^{\dagger}a$ operator. Then, by
doing a simple exercise, we find that the action of the creation
and annihilation operators, $a^{\dagger}$ and $a$, on the properly
normalized fields should read
\begin{equation}
a^{\dagger}\psi_{n}=\sqrt{n+1}\psi_{n+1},\quad 
a\psi_{n}=\sqrt{n}\psi_{n-1}.
\label{a-psi}
\end{equation}
Making use of these properties, we check that the kinetic term in
the effective action is invariant under the transformations in
Eqs.~(\ref{trans1}), (\ref{trans2}) and (\ref{trans12}). In fact,
if we had started with a more general, non-diagonal kinetic term in
the effective action [which is not forbidden by the chiral $U(1)$
symmetry], the requirement of invariance under the magnetic
translations would lead us back to the diagonal form as in
Eq.~(\ref{eff-gen}).

The invariance of the four-fermion interaction $g$ in
Eq.~(\ref{eff-gen}) under the set of transformations  in
Eqs.~(\ref{trans1}), (\ref{trans2}) and (\ref{trans12}) leads to
\begin{eqnarray}
&&\sqrt{n_1}
g\left(\begin{array}{cc}
             n_1-1 & n_2 \\
             m_1 & m_2
       \end{array}\right)
-\sqrt{n_2+1} 
g\left(\begin{array}{cc}
             n_1 & n_2+1 \\
             m_1 & m_2
       \end{array}\right)
\nonumber\\
&&+\sqrt{m_1} 
g\left(\begin{array}{cc}
             n_1 & n_2 \\
             m_1-1 & m_2
       \end{array}\right)
-\sqrt{m_2+1} 
g\left(\begin{array}{cc}
             n_1 & n_2 \\
             m_1 & m_2+1
       \end{array}\right)
=0,
\label{restrict1}\\
&&\sqrt{n_1+1} 
g\left(\begin{array}{cc}
             n_1+1 & n_2 \\
             m_1 & m_2
       \end{array}\right)
-\sqrt{n_2} 
g\left(\begin{array}{cc}
             n_1 & n_2-1 \\
             m_1 & m_2
       \end{array}\right)
\nonumber\\
&&+\sqrt{m_1+1} 
g\left(\begin{array}{cc}
             n_1 & n_2 \\
             m_1+1 & m_2
       \end{array}\right)
-\sqrt{m_2} 
g\left(\begin{array}{cc}
             n_1 & n_2 \\
             m_1 & m_2-1
       \end{array}\right)
=0,
\label{restrict2} \\
&&g\left(\begin{array}{cc}
             n_1 & n_2 \\
             m_1 & m_2
       \end{array}\right)=0, 
\quad \mbox{unless} \quad
n_1-n_2+m_1-m_2=0.
\label{nnmm}
\end{eqnarray}

\section{Calculation of the two-loop $\beta$-function}
\label{AppB}

To derive the two-loop $\beta$ function of the effective spinor
theory in Eq.~(\ref{eff-gen}), we apply the method of
Ref.~\cite{ZJ} that was used for the $\varphi^4$ scalar theory
in $d=4-\varepsilon$ dimensions. In dimensional regularization 
(with $D=2+2\varepsilon$), our renormalized Lagrangian density 
reads 
\begin{eqnarray}
&&{\cal L}_{NJL} =\sum_{n=0}^{\infty}  \left(
 Z^{(L)}_{n}\psi^{(L)*}_{n} i \partial_{+}\psi^{(L)}_{n}
+Z^{(R)}_{n}\psi^{(R)*}_{n} i \partial_{-}\psi^{(R)}_{n}
\right)
+\sum_{n_1 , n_2, m_1, m_2=0}^{\infty}\mu^{-2\varepsilon} 
G\left(\begin{array}{cc}
             n_1 & n_2 \\
             m_1 & m_2   
       \end{array}\right)
\psi^{(L)*}_{n_1} \psi^{(R)}_{n_2}
\psi^{(R)*}_{m_1} \psi^{(L)}_{m_2}\nonumber\\
&&=\sum_{n=0}^{\infty} \left(
\psi^{(L)*}_{n} i \partial_{+}\psi^{(L)}_{n}
+\psi^{(R)*}_{n} i \partial_{-}\psi^{(R)}_{n}\right)
+\sum_{n_1 , n_2, m_1, m_2=0}^{\infty}\mu^{-2\varepsilon} 
g\left(\begin{array}{cc}
             n_1 & n_2 \\
             m_1 & m_2   
       \end{array}\right)
\psi^{(L)*}_{n_1} \psi^{(R)}_{n_2}
\psi^{(R)*}_{m_1} \psi^{(L)}_{m_2}
+(\dots),
\label{ren-lag}
\end{eqnarray}
where the coupling $G$ includes the coupling constant
renormalization, $G=Z_{g}g$, and the ellipsis denotes the
counterterms. We remind the reader that the four-index coupling $g$
is non-zero only for $n_1-n_2+m_1-m_2=0$. This means that the
four-fermion term in Eq.~(\ref{ren-lag}) contains the sum only over
three indices (say, $n_1$, $n_2$, and $m_1$), while the fourth one
($m_2$) is superfluous.

Before proceeding with the actual calculation of the
$\beta$ function, we need to specify how to handle the infrared
divergences that appear in the calculation of the Feynman diagrams
\cite{ZJ,Gr}. Such divergences usually come from the propagators of
the massless fermions. If treated improperly, they could easily
obscure the calculation of the relevant diagrams and eventually lead
to a wrong result. To avoid the problem, in what follows, we modify
the infrared region by changing the fermion propagators as
follows:  
\begin{mathletters}
\begin{eqnarray}
S^{(L)}(p)=\frac{p_{-}}{p^2}\to \frac{p_{-}}{p^2-m^2},
\label{mod-prop-a}\\
S^{(R)}(p)=\frac{p_{+}}{p^2}\to \frac{p_{+}}{p^2-m^2}.
\label{mod-prop-b}
\end{eqnarray}
\end{mathletters}
This infrared regularization procedure respects all the symmetries
of the model and does not change the ultraviolet region.

Now, let us calculate the $\beta$ function. First of all, we recall
that the relation between the bare and the renormalized couplings,
in the dimensional regularization ($D=2+2\varepsilon$), reads
\begin{equation}
g_{0}\left(\begin{array}{cc}
             N_1 & N_2 \\
             M_1 & M_2   
       \end{array}\right)
=\frac{\mu^{-2\varepsilon} }
{\sqrt{Z^{(L)}_{N_1} Z^{(R)}_{M_1} 
       Z^{(R)}_{N_2} Z^{(L)}_{M_2}}}
G\left(\begin{array}{cc}
             N_1 & N_2 \\
             M_1 & M_2   
       \end{array}\right).
\end{equation}
This is going to be used in the definition of the $\beta$ function.
In calculation, we impose the following renormalization conditions:
\begin{eqnarray}
\left.\frac{\partial}{\partial p_{-}} 
\Gamma^{(2R)}_{NN}\right|_{p=0}=
\left.\frac{\partial}{\partial p_{+}} 
\Gamma^{(2L)}_{NN}\right|_{p=0}=1,\\
\left.\Gamma^{(4)}\left(\begin{array}{cc}
             N_1 & N_2 \\
             M_1 & M_2   
       \end{array}\right)\right|_{p=0}
=g\left(\begin{array}{cc}
             N_1 & N_2 \\
             M_1 & M_2   
       \end{array}\right).
\label{ren-con}
\end{eqnarray}
Note that the introduction of the effective infrared cutoff $m$
in the fermion propagators earlier allows us to use the
renormalization point at $p=0$.

\begin{figure}
\epsfbox{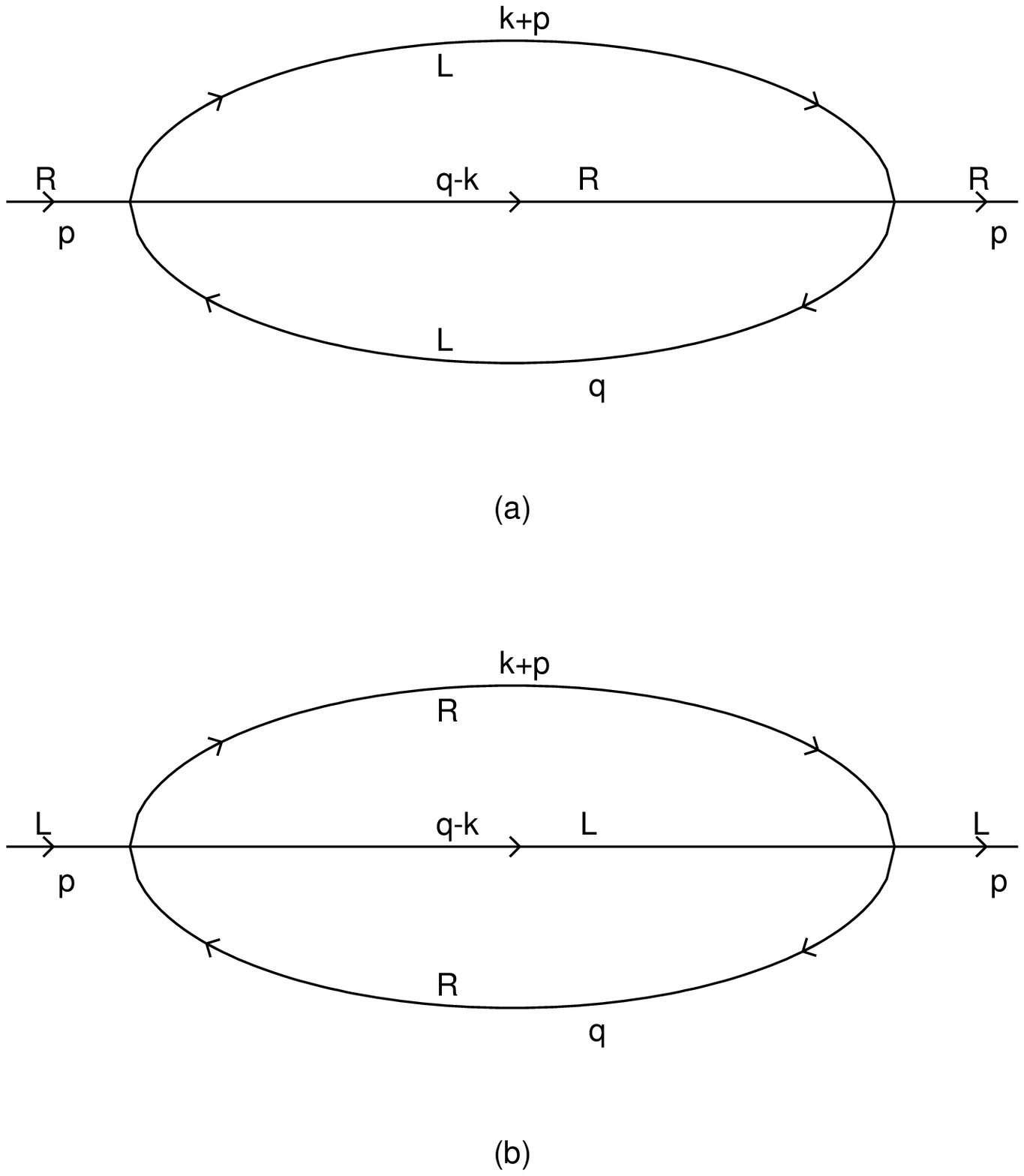}
\caption{Diagrams contributing to $\Gamma^{(2R)}$ and 
$\Gamma^{(2L)}$, respectively.}
\label{fig1}
\end{figure}

The Feynman diagrams of the relevant contributions to the two-point
functions at two-loop order are given in Figs.~\ref{fig1}a and
\ref{fig1}b. By extracting the divergent (of order $1/\varepsilon$)
terms of these two-loop corrections, we  arrive at the equations
\begin{eqnarray}
&&\left.\frac{\partial}{\partial p_{-}} 
\Gamma^{(2R)}_{NN}\right|_{p=0}
=Z^{(R)}_{N}+\frac{c^{(r)}}{(4\pi)^2\varepsilon}\sum_{k_1,k_2}
G\left(\begin{array}{cc}
             k_1 & k_2+N \\
             N & k_1-k_2     
       \end{array}\right)
G\left(\begin{array}{cc}
             k_1-k_2 & N \\
             k_2+N & k_1     
       \end{array}\right),
\label{gam-2R}\\
&&\left.\frac{\partial}{\partial p_{+}} 
\Gamma^{(2L)}_{NN}\right|_{p=0}
=Z^{(L)}_{N}+\frac{c^{(l)}}{(4\pi)^2\varepsilon}\sum_{k_1,k_2}
G\left(\begin{array}{cc}
             N & k_2+N \\
             k_1 &  k_1-k_2    
       \end{array}\right)
G\left(\begin{array}{cc}
             k_1-k_2  & k_1 \\
             k_2+N & N    
       \end{array}\right), 
\label{gam-2L}
\end{eqnarray}
where, by definition,
\begin{eqnarray}
(\mbox{Fig.~}\ref{fig1}\mbox{a}) \to \frac{c^{(r)}}
{(4\pi)^2\varepsilon}+O(1)
&\equiv&-\frac{1}{\mu^{2D-4}} 
\left.\frac{\partial}{\partial p_{-}}
\int\frac{d^{D}k d^{D}q}{(2\pi)^{2D}} 
\frac{(k+p)_{-} (q-k)_{+} q_{-}}
{[(k+p)^2-m^2][(q-k)^2-m^2](q^2-m^2)}\right|_{p=0}
\nonumber\\ &=&-\frac{1}{(4\pi)^2\varepsilon}+O(1),
\label{def-c_r} \\
(\mbox{Fig.~}\ref{fig1}\mbox{b}) \to \frac{c^{(l)}}
{(4\pi)^2\varepsilon}+O(1)
&\equiv&-\frac{1}{\mu^{2D-4}}
\left.\frac{\partial}{\partial p_{+}}
\int\frac{d^{D}k d^{D}q}{(2\pi)^{2D}} 
\frac{(k+p)_{+} (q-k)_{-} q_{+}}
{[(k+p)^2-m^2][(q-k)^2-m^2](q^2-m^2)}\right|_{p=0}
\nonumber\\ &=&-\frac{1}{(4\pi)^2\varepsilon}+O(1).
\label{def-c_l} 
\end{eqnarray}
Thus, we see that $c^{(r)}=c^{(l)}=-1$. 

\begin{figure}
\epsfbox{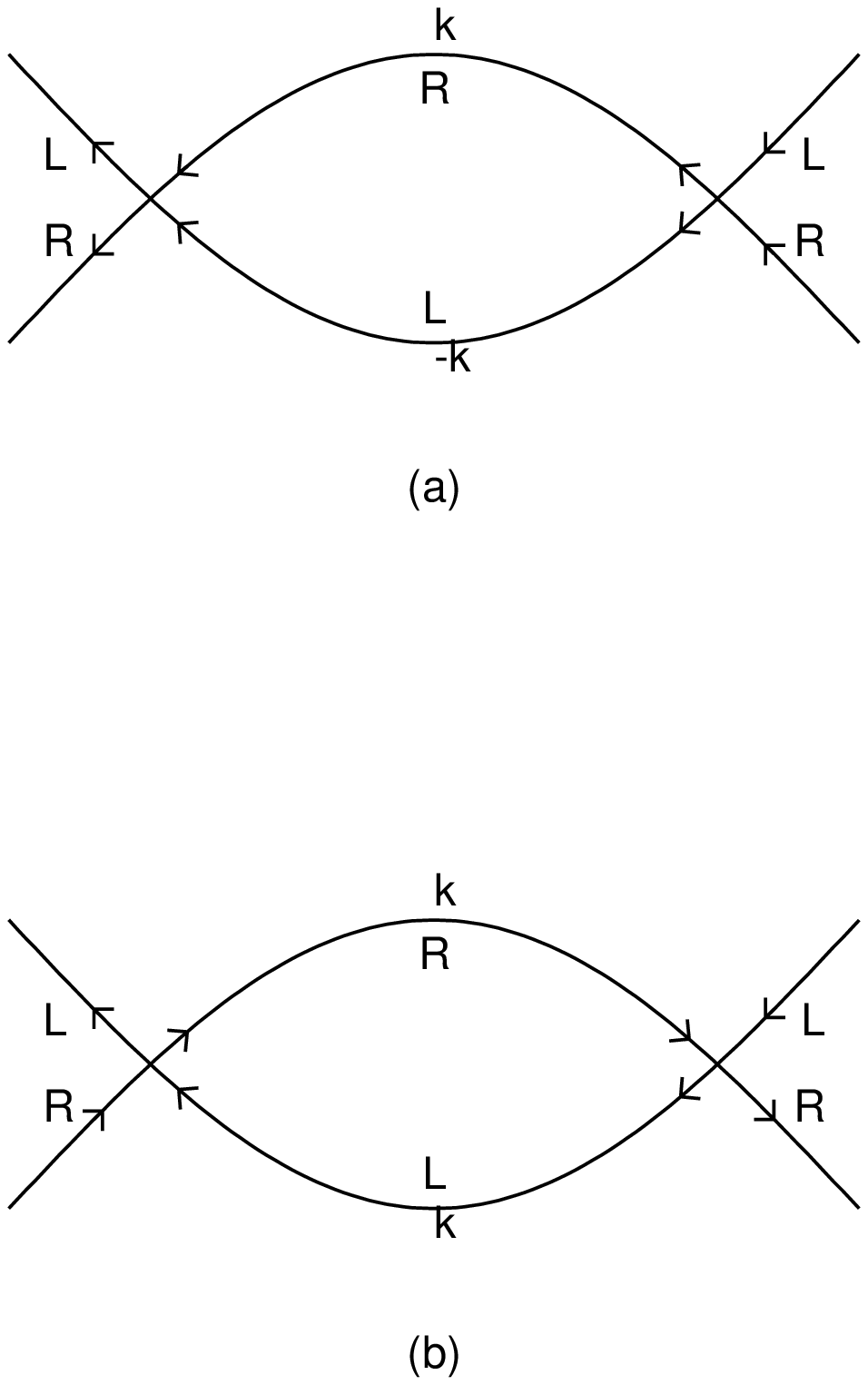}
\caption{Diagrams contributing to $\Gamma^{(4)}$ at one loop which
determine $a_0$ and $a_1$.}
\label{fig2}
\end{figure}

\begin{figure}
\epsfbox{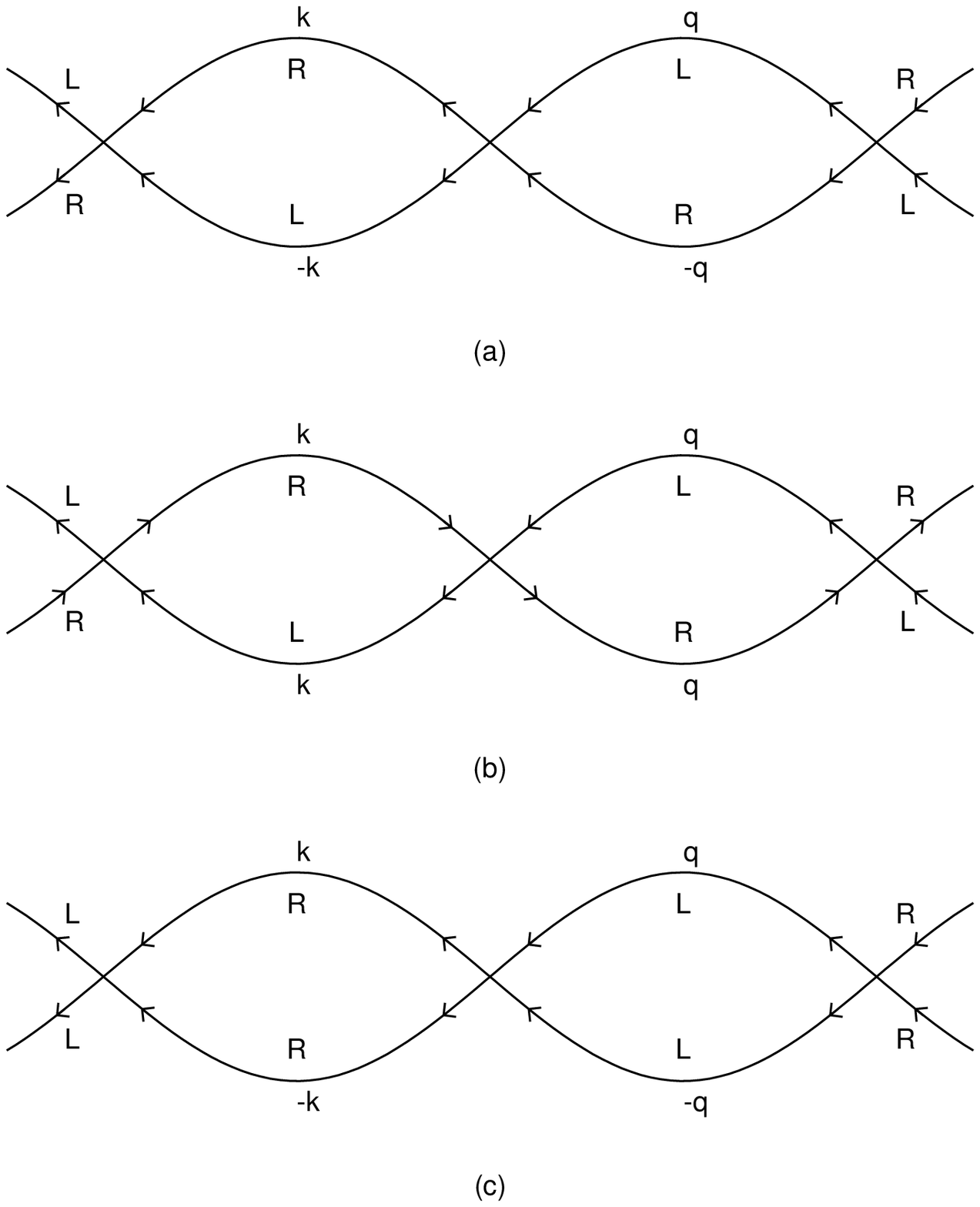}
\caption{Diagrams contributing to $\Gamma^{(4)}$ at two loops which
determine $b_0$ and $b_1$. The diagram (c) is finite.}
\label{fig3}
\end{figure}

\begin{figure}
\epsfbox{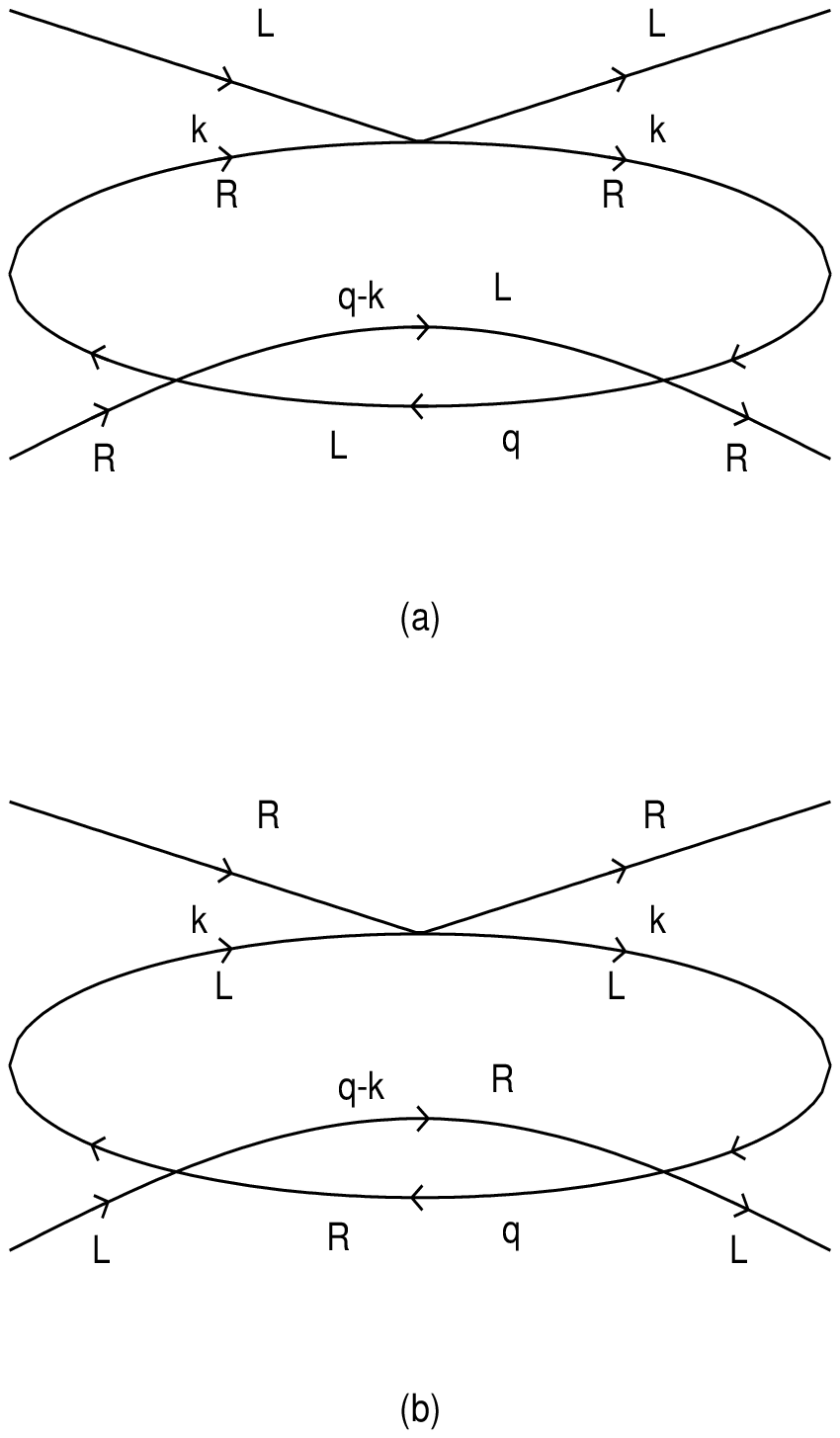}
\caption{Diagrams contributing to $\Gamma^{(4)}$ at two loops which
determine $c_0$.}
\label{fig4}
\end{figure}

\begin{figure}
\hbox{
\epsfbox{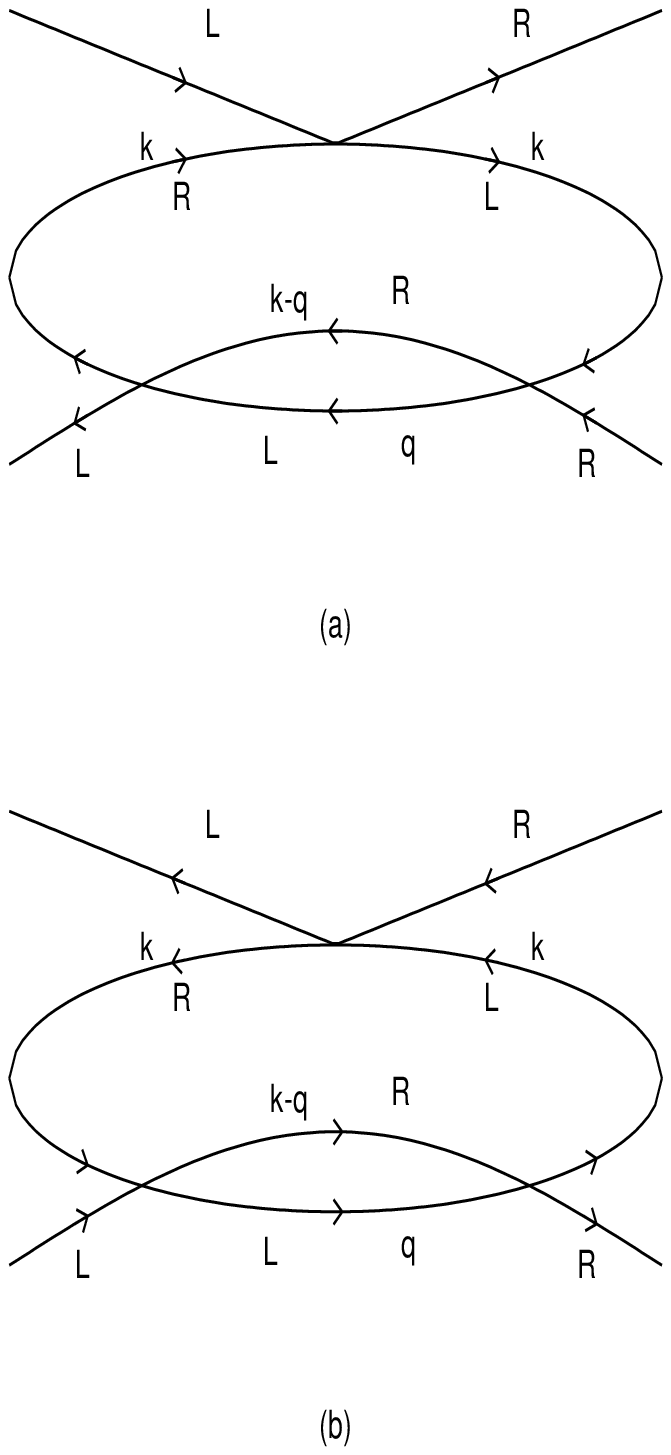}
\epsfbox{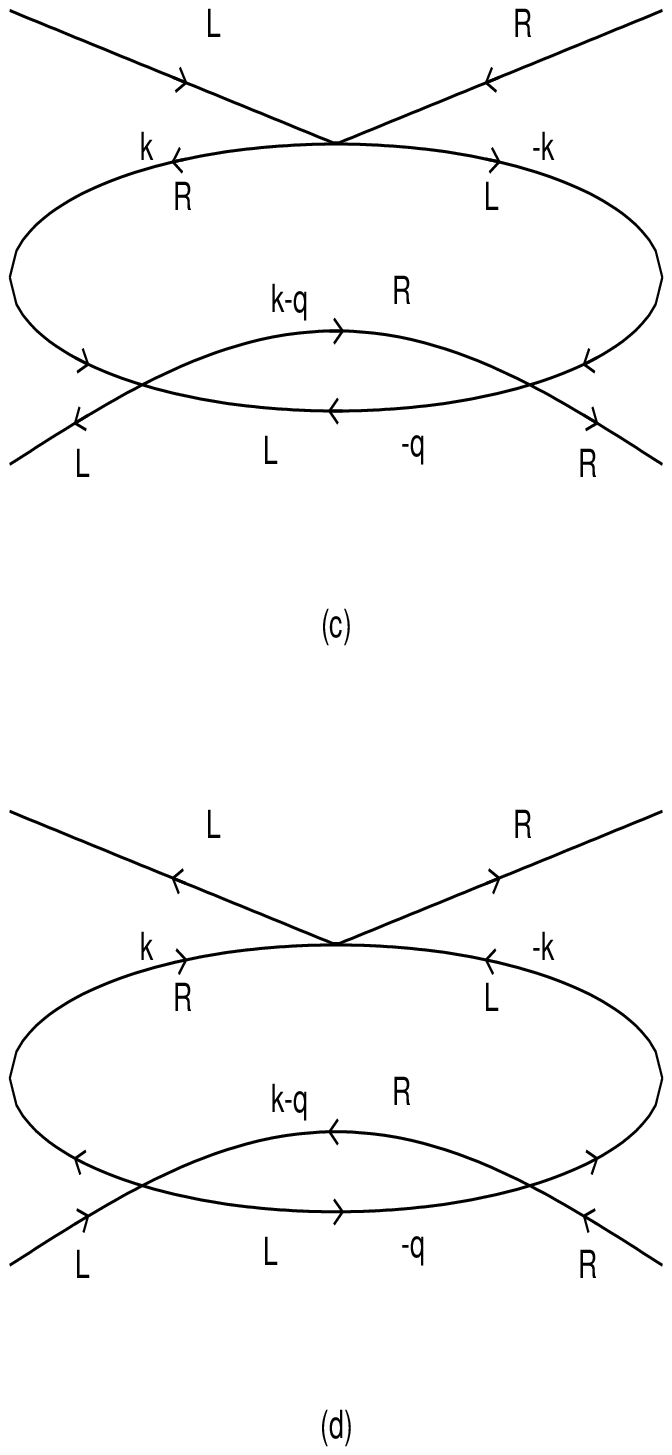} }
\caption{Diagrams contributing to $\Gamma^{(4)}$ at two loops which
determine $d_0$ and $d_1$.}
\label{fig5}
\end{figure}

In a similar way, the perturbative expansion for the four-point 
function reads 
\begin{eqnarray}
&&\left.\Gamma^{(4)}\left(\begin{array}{cc}
             N_1 & N_2 \\
             M_1 & M_2   
       \end{array}\right)\right|_{p=0}
=G\left(\begin{array}{cc}
             N_1 & N_2 \\
             M_1 & M_2   
       \end{array}\right) 
\nonumber\\
+&& \frac{a_0+a_1\varepsilon}{4\pi\varepsilon}\sum_{k_1}\left[
G\left(\begin{array}{cc}
             N_1 & N_2 \\
             k_1 & N_1-N_2+k_1     
       \end{array}\right)
G\left(\begin{array}{cc}
             N_1-N_2+k_1 & k_1 \\
             M_1 & M_2     
       \end{array}\right)\right.\nonumber\\
&&-\left.G\left(\begin{array}{cc}
             N_1 & k_1 \\
             M_1 & N_1+M_1-k_1
       \end{array}\right)
G\left(\begin{array}{cc} 
             N_1+M_1-k_1 & N_2 \\
             k_1 & M_2     
       \end{array}\right)\right]
\nonumber\\[3mm]
+&&  \frac{b_0+b_1\varepsilon}{(4\pi\varepsilon)^2}
\sum_{k_1,k_2}\left[
G\left(\begin{array}{cc}
             N_1 & N_2 \\
             k_1 & N_1-N_2+k_1     
       \end{array}\right)
G\left(\begin{array}{cc}
             N_1-N_2+k_1 & k_1 \\
             k_2 & N_1-N_2+k_2    
       \end{array}\right)
G\left(\begin{array}{cc}
             N_1-N_2+k_2 & k_2 \\
             M_1 & M_2     
       \end{array}\right)\right.\nonumber\\
&& +\left. G\left(\begin{array}{cc}
             N_1 & k_1 \\
             M_1 & N_1+M_1-k_1     
       \end{array}\right)
G\left(\begin{array}{cc}
             N_1+M_1-k_1 & k_2 \\
             k_1 & N_1+M_1-k_2    
       \end{array}\right)
G\left(\begin{array}{cc}
             N_1+M_1-k_2 & N_2 \\
             k_2 & M_2    
       \end{array}\right)\right]
\nonumber \\[3mm]
+&& \frac{c_0}{(4\pi)^2\varepsilon}
\sum_{k_1,k_2} \left[
G\left(\begin{array}{cc}
             N_1 & k_2 \\
             k_1 & N_1+k_1-k_2    
       \end{array}\right)
G\left(\begin{array}{cc}
             N_1+k_1-k_2 & N_2 \\
             M_1& M_2+k_1-k_2     
       \end{array}\right)
G\left(\begin{array}{cc}
             M_2+k_1-k_2 & k_1 \\
             k_2 & M_2    
       \end{array}\right)\right.
\nonumber\\
&&+\left.G\left(\begin{array}{cc}
             k_1 & k_2 \\
             M_1 & M_1+k_1-k_2     
       \end{array}\right)
G\left(\begin{array}{cc}
             N_1 & N_2-M_1+k_2 \\
             k_2 & M_2    
       \end{array}\right)
G\left(\begin{array}{cc}
             M_1+k_1-k_2 & N_2 \\
             N_2-M_1+k_2 & k_1    
       \end{array}\right) \right]
\nonumber\\[3mm]
+&& \frac{d_0+d_1\varepsilon}{(4\pi\varepsilon)^2}
\sum_{k_1,k_2} \left[
G\left(\begin{array}{cc}
             N_1 & k_2 \\
             k_1 & N_1+k_1-k_2     
       \end{array}\right)
G\left(\begin{array}{cc}
             N_1-N_2+k_1 & k_1 \\
             M_1 & M_2     
       \end{array}\right)
G\left(\begin{array}{cc}
             N_1+k_1-k_2 & N_2 \\
             k_2 & N_1-N_2+k_1   
       \end{array}\right)\right.
\nonumber\\
&&+\left.G\left(\begin{array}{cc}
             k_1 & k_2 \\
             M_1 & k_1-k_2+M_1    
       \end{array}\right)
G\left(\begin{array}{cc}
             N_1 & N_2 \\
             N_2-N_1+k_1& k_1   
       \end{array}\right)
G\left(\begin{array}{cc}
             k_1-k_2+M_1 & N_2-N_1+k_1 \\
             k_2 & M_2     
       \end{array}\right)\right.
\nonumber\\
&&+\left.G\left(\begin{array}{cc}
             N_1 & N_1+k_1-k_2 \\
             k_1 & k_2     
       \end{array}\right)
G\left(\begin{array}{cc}
             M_1+k_2-k_1 & N_2 \\
             N_1+k_1-k_2 & M_2     
       \end{array}\right)
G\left(\begin{array}{cc}
             k_2 & k_1 \\
             M_1 & M_1+k_2-k_1     
       \end{array}\right)\right.
\nonumber\\
&&+\left. G\left(\begin{array}{cc}
             N_1+M_1-k_2 & N_2 \\
             k_1+k_2-M_2 & k_1     
       \end{array}\right)
G\left(\begin{array}{cc}
             N_1 & k_2 \\
             M_1 & N_1+M_1-k_2     
       \end{array}\right)
G\left(\begin{array}{cc}
             k_1 & k_1+k_2-M_2 \\
             k_2 & M_2     
       \end{array}\right)\right].
\label{gam-4} 
\end{eqnarray}
Note that the restriction $n_1-n_2+m_1-m_2=0$ is satisfied for
each four-index coupling that appears here. The coefficients $a_i$,
$b_i$, $c_i$  and $d_i$ are defined by the following expressions:
\begin{eqnarray}
(\mbox{Fig.~}\ref{fig2}) \to \frac{a_0+a_1\varepsilon}
{4\pi\varepsilon}+O(\varepsilon)
&\equiv&\frac{i}{\mu^{D-2}}\int\frac{d^{D}k}{(2\pi)^{D}} 
\frac{k^2}{(k^2-m^2)^2}
\nonumber\\ &=&-\frac{1}{4\pi\varepsilon}
\left(1+\varepsilon(1+\gamma)
+\varepsilon\ln\frac{m^2}{4\pi\mu^2}\right)+O(\varepsilon),
\label{def-a} \\
(\mbox{Fig.~}\ref{fig3}) \to \frac{b_0+b_1\varepsilon}
{(4\pi\varepsilon)^2}+O(1)
&\equiv&\frac{i^2}{\mu^{2D-4}}\int\frac{d^{D}k d^{D}q}
{(2\pi)^{2D}} \frac{q^2 k^2}{ (q^2-m^2)^2 (k^2-m^2)^2}
\nonumber\\ &=&\frac{1}{(4\pi\varepsilon)^2}
\left(1+2\varepsilon(1+\gamma)
+2\varepsilon\ln\frac{m^2}{4\pi\mu^2}\right)+O(1),
\label{def-b} \\
(\mbox{Fig.~}\ref{fig4}) \to \frac{c_0}{(4\pi)^2\varepsilon}
+O(\varepsilon)
&\equiv&\frac{i^2}{\mu^{2D-4}}\int\frac{d^{D}k d^{D}q}
{(2\pi)^{2D}}\frac{k^2_{+}q_{-}(q-k)_{-}}
{(k^2-m^2)^2 (q^2-m^2) [(q-k)^2-m^2]}  \nonumber\\ 
&=& -\frac{1}{2(4\pi)^2\varepsilon}+O(\varepsilon),
\label{def-c}
\end{eqnarray}
and 
\begin{eqnarray}
(\mbox{Fig.~}\ref{fig5}) \to \frac{d_0+d_1\varepsilon}
{(4\pi\varepsilon)^2}+O(1)
&\equiv&\frac{i^2}{\mu^{2D-4}}\int\frac{d^{D}k d^{D}q}
{(2\pi)^{2D}}\frac{k^2 q_{-} (k-q)_{+}}
{(k^2-m^2)^2 (q^2-m^2) [(k-q)^2-m^2]}\nonumber\\ 
&=&-\frac{1}{(4\pi\varepsilon)^2}
\left(\frac{1}{2}+\varepsilon(1+\gamma)
+\varepsilon\ln\frac{m^2}{4\pi\mu^2}\right)+O(1).
\label{def-d}
\end{eqnarray}
Here $\gamma\approx 0.577$ is the Euler constant.

From Eqs.~(\ref{def-a}) -- (\ref{def-d}), we obtain
\begin{mathletters}
\begin{eqnarray}
a_0=-1, &\qquad& a_1
=-(1+\gamma)-\ln\frac{m^2}{4\pi\mu^2},\label{a_i}\\
b_0=1, &\qquad& b_1
=2(1+\gamma)+2\ln\frac{m^2}{4\pi\mu^2},\label{b_i}\\
c_0=-\frac{1}{2}, &\qquad&~\label{c_i}\\
d_0=-\frac{1}{2}, &\qquad& d_1=-(1+\gamma)-\ln\frac{m^2}{4\pi\mu^2}.
\label{d_i}
\end{eqnarray}
\end{mathletters}

After expressing the function $G$ in terms of $g$, we arrive at
\begin{eqnarray}
&&G\left(\begin{array}{cc}
             N_1 & N_2 \\
             M_1 & M_2   
       \end{array}\right)
=g\left(\begin{array}{cc}
             N_1 & N_2 \\
             M_1 & M_2   
       \end{array}\right) 
-\frac{a_0+a_1\varepsilon}
{4\pi\varepsilon}\sum_{k_1}\left[
g\left(\begin{array}{cc}
             N_1 & N_2 \\
             k_1 & N_1-N_2+k_1     
       \end{array}\right)
g\left(\begin{array}{cc}
             N_1-N_2+k_1 & k_1 \\
             M_1 & M_2     
       \end{array}\right)\right.
\nonumber \\
&&-\left.g\left(\begin{array}{cc}
             N_1 & k_1 \\
             M_1 & N_1+M_1-k_1
       \end{array}\right)
g\left(\begin{array}{cc} 
             N_1+M_1-k_1 & N_2 \\
             k_1 & M_2     
       \end{array}\right)\right]
-\frac{b_0-2a_0^2+(b_1-4a_1a_0)\varepsilon}
{(4\pi\varepsilon)^2} 
\nonumber\\[3mm]
&&\times \sum_{k_1,k_2} \left[
g\left(\begin{array}{cc}
             N_1 & N_2 \\
             k_1 & N_1-N_2+k_1     
       \end{array}\right)
g\left(\begin{array}{cc}
             N_1-N_2+k_1 & k_1 \\
             k_2 & N_1-N_2+k_2    
       \end{array}\right)
g\left(\begin{array}{cc}
             N_1-N_2+k_2 & k_2 \\
             M_1 & M_2     
       \end{array}\right)\right.\nonumber\\
&& +\left. g\left(\begin{array}{cc}
             N_1 & k_1 \\
             M_1 & N_1+M_1-k_1     
       \end{array}\right)
g\left(\begin{array}{cc}
             N_1+M_1-k_1 & k_2 \\
             k_1 & N_1+M_1-k_2    
       \end{array}\right)
g\left(\begin{array}{cc}
             N_1+M_1-k_2 & N_2 \\
             k_2 & M_2    
       \end{array}\right)\right]
\nonumber \\[3mm]
-&& \frac{c_0}{(4\pi)^2\varepsilon}
\sum_{k_1,k_2} \left[
g\left(\begin{array}{cc}
             N_1 & k_2 \\
             k_1 & N_1+k_1-k_2    
       \end{array}\right)
g\left(\begin{array}{cc}
             N_1+k_1-k_2 & N_2 \\
             M_1& M_2+k_1-k_2     
       \end{array}\right)
g\left(\begin{array}{cc}
             M_2+k_1-k_2 & k_1 \\
             k_2 & M_2    
       \end{array}\right)\right.
\nonumber\\
&&+\left.g\left(\begin{array}{cc}
             k_1 & k_2 \\
             M_1 & M_1+k_1-k_2     
       \end{array}\right)
g\left(\begin{array}{cc}
             N_1 & N_2-M_1+k_2 \\
             k_2 & M_2    
       \end{array}\right)
g\left(\begin{array}{cc}
             M_1+k_1-k_2 & N_2 \\
             N_2-M_1+k_2 & k_1    
       \end{array}\right) \right]
\nonumber\\[3mm]
-&& \frac{d_0+a_0^2+(d_1+2a_0a_1)\varepsilon}
{(4\pi\varepsilon)^2} \sum_{k_1,k_2} \left[
g\left(\begin{array}{cc}
             N_1 & k_2 \\
             k_1 & N_1+k_1-k_2     
       \end{array}\right)
g\left(\begin{array}{cc}
             N_1-N_2+k_1 & k_1 \\
             M_1 & M_2     
       \end{array}\right)
g\left(\begin{array}{cc}
             N_1+k_1-k_2 & N_2 \\
             k_2 & N_1-N_2+k_1   
       \end{array}\right)\right.
\nonumber\\[3mm]
&&+g\left(\begin{array}{cc}
             k_1 & k_2 \\
             M_1 & k_1-k_2+M_1    
       \end{array}\right)
g\left(\begin{array}{cc}
             N_1 & N_2 \\
             N_2-N_1+k_1& k_1   
       \end{array}\right)
g\left(\begin{array}{cc}
             k_1-k_2+M_1 & N_2-N_1+k_1 \\
             k_2 & M_2     
       \end{array}\right)
\nonumber\\
&&+g\left(\begin{array}{cc}
             N_1 & N_1+k_1-k_2 \\
             k_1 & k_2     
       \end{array}\right)
g\left(\begin{array}{cc}
             M_1+k_2-k_1 & N_2 \\
             N_1+k_1-k_2 & M_2     
       \end{array}\right)
g\left(\begin{array}{cc}
             k_2 & k_1 \\
             M_1 & M_1+k_2-k_1     
       \end{array}\right)
\nonumber\\
&&+\left. g\left(\begin{array}{cc}
             N_1+M_1-k_2 & N_2 \\
             k_1+k_2-M_2 & k_1     
       \end{array}\right)
g\left(\begin{array}{cc}
             N_1 & k_2 \\
             M_1 & N_1+M_1-k_2     
       \end{array}\right)
g\left(\begin{array}{cc}
             k_1 & k_1+k_2-M_2 \\
             k_2 & M_2     
       \end{array}\right)\right].\label{G(g)}
\end{eqnarray}
The $\beta$ function is defined as follows \cite{ZJ}:
\begin{eqnarray}
\sum_{n_1,n_2,m_1,m_2}\beta(g)\left(\begin{array}{cc} 
             n_1 & n_2 \\
             m_1 & m_2   
       \end{array}\right)
\frac{\partial}{\partial
g\left(\begin{array}{cc} 
             n_1 & n_2 \\
             m_1 & m_2   
       \end{array}\right)}
\left[\frac{
G\left(\begin{array}{cc} 
             N_1 & N_2 \\
             M_1 & M_2   
       \end{array}\right)}
{\sqrt{Z^{(L)}_{N_1} Z^{(R)}_{M_1} 
       Z^{(R)}_{N_2} Z^{(L)}_{M_2}}}\right]
=2\varepsilon\frac{
G\left(\begin{array}{cc} 
             N_1 & N_2 \\
             M_1 & M_2   
       \end{array}\right)}
{\sqrt{Z^{(L)}_{N_1} Z^{(R)}_{M_1} 
       Z^{(R)}_{N_2} Z^{(L)}_{M_2}}}.
\label{def-beta}
\end{eqnarray}
By making use of this definition, we calculate the two-loop 
$\beta$ function,
\begin{eqnarray}
\beta(g)\left(\begin{array}{cc} 
             N_1 & N_2 \\
             M_1 & M_2   
       \end{array}\right)&&
= 2 \varepsilon g\left(\begin{array}{cc} 
             N_1 & N_2 \\
             M_1 & M_2  
        \end{array}\right)
+\frac{a_0}{2\pi}\sum_{k_1}\left[
g\left(\begin{array}{cc}
             N_1 & N_2 \\
             k_1 & N_1-N_2+k_1     
       \end{array}\right)
g\left(\begin{array}{cc}
             N_1-N_2+k_1 & k_1 \\
             M_1 & M_2     
       \end{array}\right)\right.
\nonumber \\
&&-\left.g\left(\begin{array}{cc}
             N_1 & k_1 \\
             M_1 & N_1+M_1-k_1
       \end{array}\right)
g\left(\begin{array}{cc} 
             N_1+M_1-k_1 & N_2 \\
             k_1 & M_2     
       \end{array}\right)\right]
\nonumber\\[3mm]
+\frac{b_1-2a_0a_1}{4\pi^2}
\sum_{k_1,k_2}&&  \left[
g\left(\begin{array}{cc}
             N_1 & N_2 \\
             k_1 & N_1-N_2+k_1     
       \end{array}\right)
g\left(\begin{array}{cc}
             N_1-N_2+k_1 & k_1 \\
             k_2 & N_1-N_2+k_2    
       \end{array}\right)
g\left(\begin{array}{cc}
             N_1-N_2+k_2 & k_2 \\
             M_1 & M_2     
       \end{array}\right)\right.\nonumber\\
&& +\left. g\left(\begin{array}{cc}
             N_1 & k_1 \\
             M_1 & N_1+M_1-k_1     
       \end{array}\right)
g\left(\begin{array}{cc}
             N_1+M_1-k_1 & k_2 \\
             k_1 & N_1+M_1-k_2    
       \end{array}\right)
g\left(\begin{array}{cc}
             N_1+M_1-k_2 & N_2 \\
             k_2 & M_2    
       \end{array}\right)\right]
\nonumber \\[3mm]
+\frac{c_0}{4\pi^2}
\sum_{k_1,k_2}&&  \left[
g\left(\begin{array}{cc}
             N_1 & k_2 \\
             k_1 & N_1+k_1-k_2    
       \end{array}\right)
g\left(\begin{array}{cc}
             N_1+k_1-k_2 & N_2 \\
             M_1& M_2+k_1-k_2     
       \end{array}\right)
g\left(\begin{array}{cc}
             M_2+k_1-k_2 & k_1 \\
             k_2 & M_2    
       \end{array}\right)\right.
\nonumber\\
&&+\left.g\left(\begin{array}{cc}
             k_1 & k_2 \\
             M_1 & M_1+k_1-k_2     
       \end{array}\right)
g\left(\begin{array}{cc}
             N_1 & N_2-M_1+k_2 \\
             k_2 & M_2    
       \end{array}\right)
g\left(\begin{array}{cc}
             M_1+k_1-k_2 & N_2 \\
             N_2-M_1+k_2 & k_1    
       \end{array}\right) \right]
\nonumber\\[3mm]
+\frac{d_1+a_0a_1}{4\pi^2}
\sum_{k_1,k_2}&&  \left[
g\left(\begin{array}{cc}
             N_1 & k_2 \\
             k_1 & N_1+k_1-k_2     
       \end{array}\right)
g\left(\begin{array}{cc}
             N_1-N_2+k_1 & k_1 \\
             M_1 & M_2     
       \end{array}\right)
g\left(\begin{array}{cc}
             N_1+k_1-k_2 & N_2 \\
             k_2 & N_1-N_2+k_1   
       \end{array}\right)\right.
\nonumber\\
&&+g\left(\begin{array}{cc}
             k_1 & k_2 \\
             M_1 & k_1-k_2+M_1    
       \end{array}\right)
g\left(\begin{array}{cc}
             N_1 & N_2 \\
             N_2-N_1+k_1& k_1   
       \end{array}\right)
g\left(\begin{array}{cc}
             k_1-k_2+M_1 & N_2-N_1+k_1 \\
             k_2 & M_2     
       \end{array}\right)
\nonumber\\
&&+g\left(\begin{array}{cc}
             N_1 & N_1+k_1-k_2 \\
             k_1 & k_2     
       \end{array}\right)
g\left(\begin{array}{cc}
             M_1+k_2-k_1 & N_2 \\
             N_1+k_1-k_2 & M_2     
       \end{array}\right)
g\left(\begin{array}{cc}
             k_2 & k_1 \\
             M_1 & M_1+k_2-k_1     
       \end{array}\right)
\nonumber\\
&&+\left. g\left(\begin{array}{cc}
             N_1+M_1-k_2 & N_2 \\
             k_1+k_2-M_2 & k_1     
       \end{array}\right)
g\left(\begin{array}{cc}
             N_1 & k_2 \\
             M_1 & N_1+M_1-k_2     
       \end{array}\right)
g\left(\begin{array}{cc}
             k_1 & k_1+k_2-M_2 \\
             k_2 & M_2     
       \end{array}\right)\right]\nonumber\\[3mm]
+\frac{2}{(4\pi)^2}
 g\left(\begin{array}{cc}
             N_1 & N_2 \\
             M_1 & M_2     
       \end{array}\right)
&&\sum_{k_1,k_2} \left[
g\left(\begin{array}{cc}
             N_1 & k_2+N_1 \\
             k_1 & k_1-k_2    
       \end{array}\right)
g\left(\begin{array}{cc}
             k_1-k_2  & k_1 \\
             k_2+N_1 & N_1    
       \end{array}\right)
+g\left(\begin{array}{cc}
             k_1 & k_2+N_2 \\
             N_2 & k_1-k_2     
       \end{array}\right)
g\left(\begin{array}{cc}
             k_1-k_2 & N_2 \\
             k_2+N_2 & k_1     
       \end{array}\right)\right.\nonumber\\
&&+\left.g\left(\begin{array}{cc}
             k_1 & k_2+M_1 \\
             M_1 & k_1-k_2     
       \end{array}\right)
g\left(\begin{array}{cc}
             k_1-k_2 & M_1 \\
             k_2+M_1 & k_1     
       \end{array}\right)
+g\left(\begin{array}{cc}
             M_2 & k_2+M_2 \\
             k_1 & k_1-k_2    
       \end{array}\right)
g\left(\begin{array}{cc}
             k_1-k_2 & k_1 \\
             k_2+M_2 & M_2    
       \end{array}\right)\right], \label{2-loop}
\end{eqnarray}
where we already used the fact that $a_0^2=b_0=-2d_0$ which, by the
way, is the necessary condition for the renormalizability of the
model. Note that the last term in Eq.~(\ref{2-loop})  appears due
to the renormalization of the two-point function. After taking into
account the values of constants in Eqs.~(\ref{a_i}) --
(\ref{d_i}), we arrive at our final result for the $\beta$
function: 
\begin{eqnarray} 
\beta(g)\left(\begin{array}{cc} 
             N_1 & N_2 \\
             M_1 & M_2   
       \end{array}\right)&&
= 2 \varepsilon g\left(\begin{array}{cc} 
             N_1 & N_2 \\
             M_1 & M_2  
        \end{array}\right)
-\frac{1}{2\pi}\sum_{k_1}\left[
g\left(\begin{array}{cc}
             N_1 & N_2 \\
             k_1 & N_1-N_2+k_1     
       \end{array}\right)
g\left(\begin{array}{cc}
             N_1-N_2+k_1 & k_1 \\
             M_1 & M_2     
       \end{array}\right)\right.
\nonumber\\
&&-\left.g\left(\begin{array}{cc}
             N_1 & k_1 \\
             M_1 & N_1+M_1-k_1
       \end{array}\right)
g\left(\begin{array}{cc} 
             N_1+M_1-k_1 & N_2 \\
             k_1 & M_2     
       \end{array}\right)\right]
\nonumber\\[3mm]
-\frac{1}{8\pi^2}
\sum_{k_1,k_2}&&  \left[
g\left(\begin{array}{cc}
             N_1 & k_2 \\
             k_1 & N_1+k_1-k_2    
       \end{array}\right)
g\left(\begin{array}{cc}
             N_1+k_1-k_2 & N_2 \\
             M_1& M_2+k_1-k_2     
       \end{array}\right)
g\left(\begin{array}{cc}
             M_2+k_1-k_2 & k_1 \\
             k_2 & M_2    
       \end{array}\right)\right.
\nonumber\\
&&+\left.g\left(\begin{array}{cc}
             k_1 & k_2 \\
             M_1 & M_1+k_1-k_2     
       \end{array}\right)
g\left(\begin{array}{cc}
             N_1 & N_2-M_1+k_2 \\
             k_2 & M_2    
       \end{array}\right)
g\left(\begin{array}{cc}
             M_1+k_1-k_2 & N_2 \\
             N_2-M_1+k_2 & k_1    
       \end{array}\right) \right]
\nonumber\\[3mm]
+\frac{2}{(4\pi)^2}
 g\left(\begin{array}{cc}
             N_1 & N_2 \\
             M_1 & M_2     
       \end{array}\right)
&&\sum_{k_1,k_2} \left[
g\left(\begin{array}{cc}
             N_1 & k_2+N_1 \\
             k_1 & k_1-k_2    
       \end{array}\right)
g\left(\begin{array}{cc}
             k_1-k_2  & k_1 \\
             k_2+N_1 & N_1    
       \end{array}\right)
+g\left(\begin{array}{cc}
             k_1 & k_2+N_2 \\
             N_2 & k_1-k_2     
       \end{array}\right)
g\left(\begin{array}{cc}
             k_1-k_2 & N_2 \\
             k_2+N_2 & k_1     
       \end{array}\right)\right.\nonumber\\
&&+\left.g\left(\begin{array}{cc}
             k_1 & k_2+M_1 \\
             M_1 & k_1-k_2     
       \end{array}\right)
g\left(\begin{array}{cc}
             k_1-k_2 & M_1 \\
             k_2+M_1 & k_1     
       \end{array}\right)
+g\left(\begin{array}{cc}
             M_2 & k_2+M_2 \\
             k_1 & k_1-k_2    
       \end{array}\right)
g\left(\begin{array}{cc}
             k_1-k_2 & k_1 \\
             k_2+M_2 & M_2    
       \end{array}\right)\right].
\label{beta-2-loop} 
\end{eqnarray}
 
\section{NJL Couplings contain maximally symmetric ones}
\label{AppC}

While having less symmetry, the coupling in Eq.~(\ref{lll-int})
still could contain contributions of highly symmetric solutions as
in Eqs.~(\ref{UnUn}) and (\ref{Uv}). In order to extract such
contributions, we introduce the projection operators to the
corresponding subspaces in the space of couplings, 
\begin{eqnarray}
P^{(LR)}\left[\cdots\right]&=&\lim_{N\to\infty} I^{(LR)} 
\frac{N\mbox{Tr}^{(LR)}\left[\cdots\right]
-\mbox{Tr}^{(V)}\left[\cdots\right]}{ N(N^2-1) }, \label{pr-LR} \\
P^{(V)}\left[\cdots\right]&=&\lim_{N\to\infty} I^{(V)}
\frac{N\mbox{Tr}^{(V)}\left[\cdots\right]
-\mbox{Tr}^{(LR)}\left[\cdots\right] }{ N(N^2-1) },
\label{pr-V} 
\end{eqnarray}
where, by definition,
\begin{eqnarray}
I^{(LR)}\left(\begin{array}{cc}
             n_1 & n_2 \\
             m_1 & m_2
       \end{array}\right)
&=&\delta_{n_1,m_2}\delta_{n_2,m_1},\\
\mbox{Tr}^{(LR)}\left[g\left(\begin{array}{cc}
             n_1 & n_2 \\
             m_1 & m_2
       \end{array}\right)\right]
&=&\sum_{n,m} g\left(\begin{array}{cc}
             n & m \\
             m & n
       \end{array}\right), \\
I^{(V)}\left(\begin{array}{cc}
             n_1 & n_2 \\
             m_1 & m_2
       \end{array}\right)
&=&\delta_{n_1,n_2}\delta_{m_1,m_2},\\
\mbox{Tr}^{(V)}\left[g\left(\begin{array}{cc}
             n_1 & n_2 \\
             m_1 & m_2
       \end{array}\right)\right]
&=&\sum_{n,m}g\left(\begin{array}{cc}
             n & n \\
             m & m
       \end{array}\right).
\end{eqnarray}
By applying the projection operators in Eqs.~(\ref{pr-LR}) and 
(\ref{pr-V}) to the coupling in Eq.~(\ref{lll-int}), we easily
extract the symmetric contributions,
\begin{mathletters}
\begin{eqnarray}
P^{(LR)}\left[g\left(\begin{array}{cc}
             n_1 & n_2 \\
             m_1 & m_2
\end{array}\right)\right]&=&\frac{2G}{N+1} I^{(LR)}, 
\label{gLLL-LR}\\
P^{(V)}\left[g\left(\begin{array}{cc}
             n_1 & n_2 \\
             m_1 & m_2
\end{array}\right)\right]&=&\frac{2G}{N+1} I^{(V)}.
\label{gLLL-V}
\end{eqnarray}
\end{mathletters}

\newpage

\end{document}